\documentclass[10pt,letterpaper,aps,onecolumn,superscriptaddress,floatfix,notitlepage,nofootinbib]{revtex4-1}
\usepackage[latin1]{inputenc}
\usepackage{natbib,color}
\usepackage{amsmath}
\usepackage{amsfonts}
\usepackage{amssymb}
\usepackage{bm}
\usepackage{graphicx}
\usepackage{graphicx}
\begin{document}
	\title{Time reversal symmetry of generalized quantum measurements with past and future boundary conditions} \author{Sreenath K. Manikandan}
	\email{skizhakk@ur.rochester.edu}
	\affiliation{Department of Physics and Astronomy, University of Rochester, Rochester, NY 14627, USA}
	\affiliation{Center for Coherence and Quantum Optics, University of Rochester, Rochester, NY 14627, USA}   
	\author{Andrew N. Jordan}
	\email{jordan@pas.rochester.edu}
	\affiliation{Department of Physics and Astronomy, University of Rochester, Rochester, NY 14627, USA}
	\affiliation{Center for Coherence and Quantum Optics, University of Rochester, Rochester, NY 14627, USA}    
	\affiliation{Center for Quantum Studies, Chapman University, Orange, CA, USA, 92866}
	\date{\today}
	
	\begin{abstract}
We expand the time reversal symmetry arguments of quantum mechanics, originally proposed by Wigner in the context of unitary dynamics, to contain situations including generalized measurements for monitored quantum systems. We propose a scheme to derive the time reversed measurement operators by considering the Schr\"{o}dinger picture dynamics of a qubit coupled to a measuring device, and show that the time reversed measurement operators form a Positive Operator Valued Measure (POVM) set. We present three particular examples to illustrate time reversal of measurement  operators  : (1) the Gaussian spin measurement, (2) a dichotomous POVM for spin, and (3) the measurement of qubit fluorescence. We then propose a general rule to   unravel   any rank two qubit measurement, and show that the  backward   dynamics obeys the retrodicted equations of the forward dynamics starting from the time reversed final state. We demonstrate the time reversal invariance of dynamical equations using the example of qubit fluorescence. We also generalize the discussion of a statistical arrow of time for continuous quantum measurements introduced by Dressel et al. [Phys. Rev. Lett. 119, 220507 (2017)]: we show that the backward probabilities can be computed from a process similar to retrodiction from the time reversed final state, and extend the definition of an arrow of time to ensembles prepared with pre- and post-selections, where we obtain a non-vanishing arrow of time in general. We discuss sufficient conditions for when time's arrow vanishes and show our method also captures the contributions to time's arrow due to natural physical processes like relaxation of an atom to its ground state. As a special case, we recover the time reversibility of the weak value as its complex conjugate using our method, and discuss how our conclusions differ from the time-symmetry argument of Aharonov-Bergmann-Lebowitz (ABL) rule.
	\end{abstract}
	
	\maketitle
\section{Introduction}
Although most~\footnote{A non-vanishing electric dipole moment of any elementary particle is an interesting example of a microscopically broken time reversal symmetry~\cite{shapiro1968electric,purcell1950possibility,fortson2003search}.} of the microscopic laws of physics are invariant under a suitable time reversal symmetry operation, there seems to exist a preferred ordering in which events are more likely to happen than otherwise, in the macroscopic world. This is true for a box of an ideal gas -- where one can practically keep track of the dynamics of every single molecule given their initial conditions, and knowing all the microscopic interactions, while the second law of thermodynamics dictates that the gas molecules within the box re-distribute themselves and evolve towards a final state where the entropy is a maximum~\cite{callen1985thermodynamics} -- and in cosmology, from the observation of an expanding universe~\cite{hawking1985arrow}. These apparent asymmetrical notions of time are also manifest in our everyday experiences as conscious observers; According to Wheeler, our notion of a past corresponds to experiencing a definite, informative, thus recordable set of events~\cite{zurek2003decoherence,maccone2009quantum}, and from an information theory perspective, they correspond to processes where entropy always increases or remains constant~\cite{maccone2009quantum}. Understanding how a definite arrow of time emerges from a time reversal invariant microscopic dynamics is considered to be one of the challenging questions that comprise seemingly very different disciplines of physics~\cite{hawking1985arrow,lebowitz1993boltzmann,struppa2013quantum,maccone2009quantum,coveney1991arrow,gold1962arrow}. 

Quantum theory is also invariant under a set of symmetry operations, and the time reversal symmetry in quantum mechanics was proposed as a discrete symmetry by Wigner~\cite{sakurai2017modern,wigner2012group,gottfried2013quantum,haake2013quantum}: the  Schr\"{o}dinger equation for a time reversal invariant Hamiltonian is also invariant under time reversal ($t \rightarrow -t$), if we perform a complex-conjugation of the wavefunction in the position basis~\cite{sakurai2017modern,gottfried2013quantum,haake2013quantum}. The quantum system traces a corresponding trajectory backwards in the time reversed reference frame, following the action of the time reversal operator, which we denote by using the symbol $\Theta$. Similarly, there are simple transformation rules for observables under the time reversal operation,
\begin{equation}
\Theta~\hat{x}~\Theta^{-1}=\hat{x},\hspace{1cm}\Theta~\hat{p}~\Theta^{-1}=-\hat{p},\hspace{1cm}\Theta~\hat{J}~\Theta^{-1}=-\hat{J},
\end{equation}
in analogy with the time reversal properties of the corresponding classical variables $x$, $p$, and $J$, which agree with our intuitive notions of time reversal in classical dynamics~\cite{sakurai2017modern,gottfried2013quantum,haake2013quantum}. Similar to the angular momentum observable $\hat{J}$, the spin operator in quantum mechanics, $\hat{S}$, also flips sign under time reversal~\cite{sakurai2017modern,gottfried2013quantum,haake2013quantum},
\begin{equation}
\Theta~\hat{S}~\Theta^{-1}=-\hat{S}.
\end{equation}

A natural way of probing the time symmetry of quantum systems is by introducing appropriate measurements. While naively it would seem to destroy the time reversal invariance of the quantum system -- as the conventional (strong) measurement corresponds to a many-to-one mapping of rays in the Hilbert space --   we will show that by considering generalized measurements, we can represent dynamical equations which describe the measurement process in a time reversal invariant way.     Different approaches have been made to bring in the notion of reversibility of measurement protocols earlier too, see the references~\cite{katz2008reversal,korotkov2006undoing} for specific results on superconducting/solid state qubits as examples. Similarly, Dressel et al., have studied the emergence of a statistical arrow of time, while demanding time reversal invariance for the reduced dynamical equations of motion of the quantum system (a qubit in their case), being monitored in a time-continuous manner~\cite{dressel2017arrow}. We generalize and extend their observations using the time reversal symmetry arguments of quantum mechanics, originally proposed by Wigner. In particular, we consider a single qubit interacting with a quantum measuring device in the Schr\"{o}dinger picture, and provide an extended discussion of the time reversal symmetry of quantum mechanics in this setting, using the operator sum representation (also known as Kraus representation)~\cite{sudarshan1961stochastic,jordan1961dynamical,kraus1971general,kraus1983states}.  We discuss the unraveling of an arbitrary rank two qubit measurement, and show that the backward   equations correspond to the retrodicted dynamics of the forward measurement~\cite{tan2015prediction}, starting from the time reversed final state. The special status of quantum non demolition measurements in these protocols is also discussed.

We emphasize that our approach to time reversal  invariance   in quantum measurements is quite different from similar attempts in the past~\cite{crooks2008quantum,bedingham2017time,aharonov1964time,bednorz2013noninvasiveness,bitbol1986time}, since  we necessitate time reversal invariance in the strongest form: in terms of the invariance of dynamical equations that describe the forward/backward measurement processes. This is achieved in two steps. First, we propose a scheme to time reverse the measurement operators of the forward dynamics (Sec.~\ref{weak}). For a given forward dynamics resulting from a sequence of generalized measurements, this associates a corresponding trajectory in the time reversed frame of reference. In the second step, we unravel the measurement in the time reversed reference frame, starting from the time reversed final state (backward dynamics, Sec.~\ref{Sec. TR}) and ending at the time reversed initial state. We find that the forward and the backward dynamics obey the same set of dynamical equations, demonstrating time reversal invariance in the strongest form for the measurement process.   Our approach can be compared to a quite useful method introduced by Crooks~\cite{crooks2008quantum}, where he derives the time reversed measurement operators for quantum systems from an operational point of view. His method explicitly relies on certain steady states of the quantum system being monitored to define the set of time reversed measurement operators, by requiring time symmetric probabilities of measurement outcomes with respect to the particular steady state.  While that has many operational advantages, especially when it comes to discussions regarding the thermodynamics of quantum systems~\cite{elouard2017role,elouard2016stochastic}, such operational approaches lack the insights from a true microscopic time reversal invariance due to their sensitive dependence on particular steady states of the system a priori. The advantage of our method is that we do not rely on any particular state of the system, while retaining the time reversal invariance for probabilities of measurement outcomes as a consequence of the microscopic time reversal invariance. We would also like to point out that a similar approach has been implemented to characterize entropy production along quantum trajectories using a two measurement protocol, by Manzano et al~\cite{manzano2017quantum}. In contrast, by using a continuous measurement scheme, we are able to relate the time symmetry we describe to the invariance of the governing dynamical equations in a simple way.  

We also discuss how a statistical arrow of time can still emerge within our framework, and make qualitative predictions about when the arrow of time would run backward, although the dynamical equations are time reversal invariant. Our approach is in agreement with the intuitive notion that a definite arrow of time can arise because the events happening in one particular order are more probable or likely to occur than the time reversed sequence of events happening in the reverse order, leading to the definition of a statistical arrow of time as in references~\cite{dressel2017arrow,parrondo2009entropy}. A brief and comprehensive introduction to the development of the notion of arrow of time in quantum mechanics can be found in reference~\cite{benoist2016entropy}. We also extend the definition of a statistical arrow of time introduced in~\cite{dressel2017arrow} to ensembles prepared in a time symmetric manner via pre- and post-selections and to imperfect measurements. We identify particular examples when the arrow of time would vanish in pre- and post-selected ensembles, but we stress that this happens only for special boundary states.  Generally the arrow of time will not vanish in pre- and post-selected ensembles, and we discuss how this is conceptually different from the ABL rule~\cite{aharonov1964time}, originating from a time symmetric formulation of quantum mechanics, in which the quantum arrow of time vanishes in pre- and post-selected ensembles. We also show that our arrow of time measure is able to capture the contribution to the time's arrow due to natural physical processes such as the relaxation of a qubit to its ground state in the case of qubit fluorescence. 

This  article   is organized as follows.  In Sec.~\ref{weak}, we describe the time reversal of measurement operators by considering the Schr\"{o}dinger picture dynamics of a qubit coupled to a measuring device. The unraveling of the measurement process is discussed in Sec.~\ref{Sec. TR}.  We generalize the definition of a statistical arrow of time discussed by Dressel et al.~\cite{dressel2017arrow} in Sec~\ref{arrow}, and extend it to ensembles prepared with past and future boundary conditions.   Defining the arrow of time for imperfect measurements is discussed in Appendix~\ref{imperfect}. The methods we use in this paper also allows us to study what happens to the weak value in generalized quantum measurements under time reversal, and we discuss this in Sec.~\ref{weak2}. We substantiate our methods for extending time reversal symmetry arguments to generalized measurements by considering complementary situations worked out in the appendices.  
\section{Time reversed measurement operators for generalized measurements\label{weak}}  
\begin{figure}
	\includegraphics[scale=0.5]{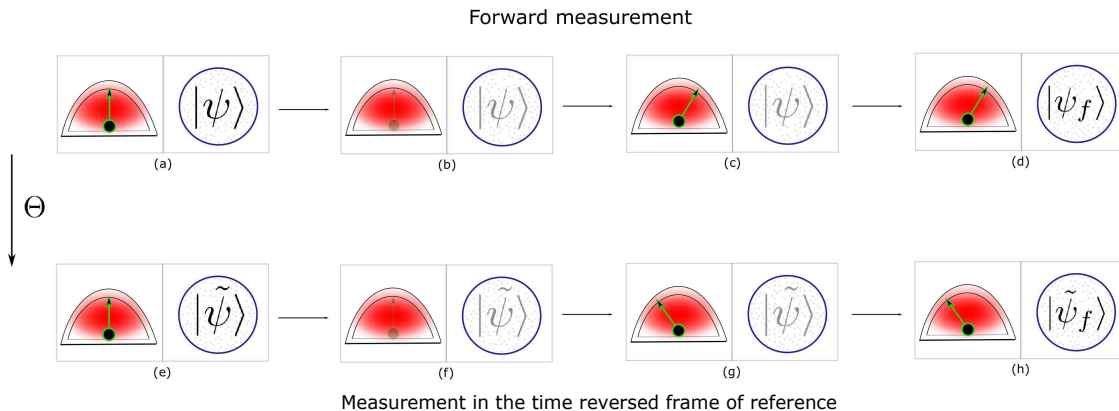}
	\caption{Here is an illustration of our general prescription to defining the time reversed measurement operators. In the forward measurement, (a) the qubit is initialized in an unknown quantum state $|\psi\rangle$. (b) The system and the measuring device evolves unitarily and becomes entangled. (c) A strong measurement on the measuring device reveals the state of the measuring device after the interaction. (d) The state of the qubit is updated, conditioned on the state of the measuring device, $|\psi_{f}\rangle\propto\hat{M}_{F}|\psi\rangle.$ In our approach to defining time reversed measurement operators, (e) first the initial state of the measuring device and the qubit are time reversed. The time reversed qubit state is $|\tilde{\psi}\rangle = \Theta|\psi\rangle.$ (f) The system and measuring device evolves under the time reversed forward unitary that entangles the qubit with the meter device. The joint state would become the time reversal of the state in (b), obtained by applying the time reversal operator $\Theta$. (g) The measuring device is projected on to the time reversed measurement state obtained in (c). If the readout was momentum of the measuring device $p$ in the forward case, we assume that we obtain the readout $-p$ in the measurement in the  time reversed frame of reference   (this scenario is worked out in Appendix~\ref{appA}). (h) The qubit state is updated conditioned on the measurement readout in step (g), $|\tilde{\psi}_{f}\rangle\propto\tilde{M}_{R}|\tilde{\psi}\rangle,$ where we assert $\tilde{M}_{R}\equiv\Theta\hat{M}_{F}\Theta^{-1}$.\label{fig0}}
\end{figure} 
 In this section, we consider three particular examples of spin and qubit fluorescence measurements, and then a generic rank two qubit measurement, to demonstrate our scheme for obtaining measurement operators in the time reversed frame of reference (time reversed measurement operators). Applying this scheme to  a sequence of measurements associates a time reversed trajectory for a given forward quantum trajectory. In section~\ref{Sec. TR}, we discuss the unraveling of the time reversed trajectory backwards, starting from the time reversed final state and ending at the time reversed initial state ( generating the ``backward trajectory"). We also show that the forward and the backward trajectories are described by the same set of dynamical equations, nevertheless, in Sec.~\ref{arrow} we discuss how a statistical arrow of time~\cite{dressel2017arrow} emerges by comparing probabilities of the forward and backward trajectories.
	
We first discuss a simple model for qubit measurements described by Gaussian measurement operators~\cite{aharonov1988result,duck1989sense}. This is a  measurement scheme well known to experimentalists; the references~\cite{vijay2012stabilizing,hacohen2016quantum,weber2014mapping,naghiloo2018information} present a variety of different paradigms where our conclusions can be investigated. Our approach to deriving the time reversed measurement operators can also be used in situations when the readout is not continuous, but discrete. To illustrate this, we consider a particular case when the measuring device is also a qubit, in Sec.~\ref{qubitmeter}. This measurement scheme has been used in rapid purification of a qubit, using feedback control~\cite{jacobs2003project}. We then study the time reversal of qubit fluorescence measurement operators in Sec.~\ref{Fl}. The references~\cite{campagne2014observing,campagne2016observing,naghiloo2017thermodynamics,naghiloo2017quantum} describe some of the contexts where fluorescence measurement has been investigated experimentally, and are potential scenarios where our unraveling techniques discussed in Sec.~\ref{Sec. TR} can be applied. We also propose a general rule to time reverse a generic rank two qubit measurement operator in Sec.~\ref{gen}. The forward and the time reversed measurement operators are also related by a probability equivalence, which is discussed in Appendix.~\ref{equivP}.   Please see Fig.~\ref{fig0} for a pictorial illustration of our approach to defining time reversed measurement operators. 
\subsection{Gaussian measurement operators for continuous spin measurement\label{SecGaus}}
 Here we review a quantum measurement model for obtaining Gaussian measurement operators~\cite{aharonov1988result,duck1989sense,barchielli1982model,caves1987quantum} for spin measurements, and obtain the corresponding operators in the time reversed reference frame.   The quantum system being measured is a spin $1/2$ particle (a qubit) and the measuring device is modeled as a continuous variable quantum system with quadratures $\hat{x}$ and $\hat{p}$. The relevant interaction between the qubit and the measuring device is described by a time reversal invariant Hamiltonian $\hat{H}_{int} $:
\begin{equation}
\hat{H}_{int}  = k~\hat{p}\hat{S}.
\end{equation}  
The Hamiltonian $\hat{H}_{int} $ is time reversal invariant since both the observables $\hat{p}$ and $\hat{S}$ are odd under time reversal. The quantum system is coupled to a measuring device via momentum operator $\hat{p}$ of the measuring device, and the readouts correspond to the measurement of the position, $\hat{x}$ of the measuring device. The complementary situation, where the readouts correspond to the momentum $\hat{p}$ of the measuring device, and the system of interest is coupled to the position $\hat{x}$ of the measuring device, is discussed in Appendix.~\ref{appA}. The derivation for Gaussian measurement operators presented here is quite similar to the approaches in references~\cite{barchielli1982model,caves1987quantum}.

We assume that the measuring device is initialized in a zero mean Gaussian quantum state,
\begin{equation}
|\psi_{m}\rangle = \frac{1}{(2\pi\delta^{2})^{1/4}}\int_{-\infty}^{\infty}dx~e^{-\frac{x^{2}}{4\delta^{2}}}~|x\rangle.
\end{equation}
We will use the notation $|\tilde{\psi}_{m}\rangle$ to indicate the time reversed quantum state, $\Theta|\psi_{m}\rangle$. We assume that the unknown quantum state of the system under measurement can be expressed in the eigenbasis of the observable $\hat{S}$,
\begin{equation}
|\psi_{s}\rangle = \sum_{n}c_{n}|s_{n}\rangle.
\end{equation}
 The joint initial state of quantum system and the measuring device is given by,
 \begin{equation}
 |\psi(0)\rangle = |\psi_{m}\rangle\otimes|\psi_{s}\rangle.
 \end{equation}
 The quantum system and the measuring device evolve unitarily for a duration $\tau$, generated by the interaction Hamiltonian $\hat{H}_{int} $, 
 \begin{eqnarray}
 |\psi(\tau)\rangle &=& e^{-ik\hat{p}\hat{S}\tau}|\psi(0)\rangle =\sum_{n'} \frac{1}{(2\pi\delta^{2})^{1/4}}\int_{-\infty}^{\infty}dx~e^{-\frac{x^{2}}{4\delta^{2}}}~e^{-ik\hat{p}s_{n'}\tau}|x\rangle\otimes |s_{n'}\rangle\langle s_{n'}||\psi_{s}\rangle\\
 &=& \sum_{n'} \frac{1}{(2\pi\delta^{2})^{1/4}}\int_{-\infty}^{\infty}dx~e^{-\frac{x^{2}}{4\delta^{2}}}|x+k\tau s_{n'}\rangle\otimes |s_{n'}\rangle\langle s_{n'}||\psi_{s}\rangle
 \\&=&\sum_{n'} \frac{1}{(2\pi\delta^{2})^{1/4}}\int_{-\infty}^{\infty}dy~e^{-\frac{(y-k\tau s_{n'})^{2}}{4\delta^{2}}}|y\rangle\otimes |s_{n'}\rangle\langle s_{n'}||\psi_{s}\rangle\\
 &=& \frac{1}{(2\pi\delta^{2})^{1/4}}\int_{-\infty}^{\infty}dy~e^{-\frac{(y-k\tau \hat{S})^{2}}{4\delta^{2}}}|y\rangle\otimes |\psi_{s}\rangle.
 \label{fwd}
  \end{eqnarray}
 The measurement is completed by projecting the measuring device onto the $y$ basis, giving the reading $\bar{y}$. The state of the quantum system change accordingly to the new (unnormalized) state,
 \begin{equation}
 \langle\bar{y}|\psi(\tau)\rangle = \frac{1}{(2\pi\delta^{2})^{1/4}}e^{-\frac{(\bar{y}-k\tau \hat{S})^{2}}{4\delta^{2}}}|\psi_{s}\rangle.
 \end{equation}  
 This defines the forward measurement operator for the measurement,
 \begin{equation}
 \hat{M}_{F}(\bar{y}) = \frac{1}{(2\pi\delta^{2})^{1/4}}e^{-\frac{(\bar{y}-k\tau \hat{S})^{2}}{4\delta^{2}}}.
 \end{equation}
 The measurement operators on the continuous variable $\bar{y}$ form a POVM set with the following completeness relation,
 \begin{equation}
 \int_{-\infty}^{\infty}d\bar{y}~\hat{M}_{F}^{\dagger}(\bar{y})\hat{M}_{F}(\bar{y}) = \hat{\mathbb{I}}.
 \end{equation}
 \subsubsection{Time reversed measurement operators}
 Now we derive the time reversed measurement operators from the time reversal symmetry arguments of quantum mechanics. We first compute the time reversal of the quantum state $|\psi(\tau)\rangle$,
 \begin{equation}
|\tilde{\psi}(\tau)\rangle = \Theta|\psi(\tau)\rangle = \Theta e^{-ik\hat{p}\hat{S}\tau}\Theta^{-1}\Theta|\psi(0)\rangle = e^{ik\hat{p}\hat{S}\tau}\Theta|\psi(0)\rangle.
 \end{equation}
Here we use the fact that under the transformation, $\hat{p}$ and $\hat{S}$ both change sign, and the expression is complex conjugated. The statement above is also a re-statement of time reversal invariance of quantum mechanics in our particular case. It implies that the time reversal of a state after a finite time $\tau$ evolving under a time reversal invariant Hamiltonian is the same as the state obtained by time reversing the initial state and evolving backward in time for the same duration $\tau$. In analogy with Eq.~\eqref{fwd}, we can re-write $|\tilde{\psi}(\tau)\rangle$, 
 \begin{eqnarray}
 |\tilde{\psi}(\tau)\rangle&=&\sum_{n'} \frac{1}{(2\pi\delta^{2})^{1/4}}\int_{-\infty}^{\infty}dx~e^{-\frac{x^{2}}{4\delta^{2}}}~e^{ik\hat{p}s_{n'}\tau}|x\rangle\otimes |s_{n'}\rangle\langle s_{n'}|\Theta|\psi_{s}\rangle\\
 &=& \sum_{n'} \frac{1}{(2\pi\delta^{2})^{1/4}}\int_{-\infty}^{\infty}dx~e^{-\frac{x^{2}}{4\delta^{2}}}|x-k\tau s_{n'}\rangle\otimes |s_{n'}\rangle\langle s_{n'}|\Theta|\psi_{s}\rangle
 \\&=&\sum_{n'}
 \frac{1}{(2\pi\delta^{2})^{1/4}}\int_{-\infty}^{\infty}dy~e^{-\frac{(y+k\tau s_{n'})^{2}}{4\delta^{2}}}|y\rangle\otimes |s_{n'}\rangle\langle s_{n'}|\Theta|\psi_{s}\rangle\\
  &=&\frac{1}{(2\pi\delta^{2})^{1/4}}\int_{-\infty}^{\infty}dy~e^{-\frac{(y+k\tau \hat{S})^{2}}{4\delta^{2}}}|y\rangle\otimes \Theta|\psi_{s}\rangle,
 \end{eqnarray}
 where we have used the time reversal invariance property of the state $|\psi_{m}\rangle$ in the position space. We now propose that the time reversed measurement operator is defined by projecting the state onto the time reversed measuring device quantum state, $|\tilde{\bar{y}}\rangle = \Theta|\bar{y}\rangle = |\bar{y}\rangle$, corresponding to the measurement readout $\bar{y}$:
  \begin{equation}
  \langle\bar{y}|\tilde{\psi}(\tau)\rangle = \frac{1}{(2\pi\delta^{2})^{1/4}}e^{-\frac{(\bar{y}+k\tau \hat{S})^{2}}{4\delta^{2}}}\Theta|\psi_{s}\rangle.
  \end{equation}  
  This defines the time reversed measurement operators $\tilde{M}_{R}(\bar{y})$ for the readout $\bar{y}$,
   \begin{equation}
   \tilde{M}_{R}(\bar{y}) = \frac{1}{(2\pi\delta^{2})^{1/4}}e^{-\frac{(\bar{y}+k\tau \hat{S})^{2}}{4\delta^{2}}}.
   \label{emmar}
   \end{equation}
    The reverse measurement operators $\tilde{M}_{R}(\bar{y})$ form a POVM obeying the following completeness relation,
    \begin{equation}
    \int_{-\infty}^{\infty}d\bar{y}~\tilde {M}_{R}^{\dagger}(\bar{y})\tilde{M}_{R}(\bar{y}) = \hat{\mathbb{I}}.
    \end{equation}
    We note that the functional form of the time reversed measurement operators are related to the corresponding time-forward measurement operators via a sign flip of the measurement record $\bar{y}$: 
     \begin{equation}
     \tilde{M}_{R}(\bar{y}) = \hat{M}_{F}(-\bar{y}),\label{Gsign}
     \end{equation}
     and we make the following assertion for the Gaussian measurement operators under time reversal operation, $\Theta$,
     \begin{equation}
     \Theta\hat{M}_{F}(\bar{y})\Theta^{-1}\equiv\tilde{M}_{R}(\bar{y}).
     \end{equation}
     This result is equivalent to the result obtained by Dressel et al., in~\cite{dressel2017arrow}. The advantage of our method is that we explicitly derive the time reversed measurement operators for a qubit and Gaussian measurement, from first principle calculations involving the Schr\"{o}dinger picture dynamics of a qubit coupled to a measuring device, and show that the time reversed measurement operators thus derived also satisfy a completeness relation similar to their forward counterparts. The case we discussed here is special, since the interaction Hamiltonian $\hat{H}_{int} $ itself is invariant under the time reversal symmetry operation. In the appendix, we show that this is not a strong requirement, and our approach can be extended to more general cases where Hamiltonian is appropriately modified under the time reversal operation, dictated by time reversal symmetry arguments of quantum mechanics.      
     \subsection{A dichotomous POVM for spin measurement\label{qubitmeter}}     
     We now consider a rather simpler model that generates the two outcome POVM used in continuously monitored qubits~\cite{jacobs2003project}, and derive the time reversed measurement operators in this discrete example using our approach. We consider a qubit interacting with another qubit (which we call the measuring device), via the controlled-NOT (c-NOT) unitary operation~\cite{nielsen2010quantum}:
     \begin{equation}
     \hat{U}_{\text{c-NOT}} = \mathbb{I}^{t}\otimes|0\rangle\langle 0|^{c}+\sigma_{x}^{t}\otimes|1\rangle\langle 1|^{c},
     \end{equation}
    where $c$ and $t$ label control and target qubit respectively. We suppress the labels in the following discussion for brevity.  Let us assume that the unknown state of the qubit being monitored is in the state $|\psi_{s}\rangle$, and the measuring device (qubit again) is initialized in the state $|\psi_{m}\rangle$,
     \begin{equation}
     |\psi_{m}\rangle = \sqrt{\gamma}|0_{m}\rangle +\sqrt{1-\gamma}|1_{m}\rangle.
     \end{equation}
     The combined state of the system and the measuring device, $|\psi_{f}\rangle$ after the c-NOT operation, where the measuring device is considered to be the target bit and the system being monitored to be the control, is given by,
     \begin{equation}
     |\psi_{f}\rangle
     =(\sqrt{\gamma}|0_{m}\rangle +\sqrt{1-\gamma}|1_{m}\rangle)\otimes|0\rangle\langle 0||\psi_{s}\rangle+(\sqrt{\gamma}|1_{m}\rangle +\sqrt{1-\gamma}|0_{m}\rangle)\otimes|1\rangle\langle 1||\psi_{s}\rangle.
     \end{equation}
     The measurement is completed by projecting the measuring device in the computational basis $\{|0_{m}\rangle,~|1\rangle_{m}\}$, which defines the following forward measurement operators,
     \begin{eqnarray}
     \langle 0_{m} |\psi_{f}\rangle &=& \big[\sqrt{\gamma}|0\rangle\langle 0|+\sqrt{1-\gamma}|1\rangle\langle 1|\big]|\psi_{s}\rangle = \hat{M}_{F}(0)|\psi_{s}\rangle,~\text{and}\\
     \langle 1_{m} |\psi_{f}\rangle &=& \big[\sqrt{1-\gamma}|0\rangle\langle 0|+\sqrt{\gamma}|1\rangle\langle 1|\big]|\psi_{s}\rangle = \hat{M}_{F}(1)|\psi_{s}\rangle.~~
     \end{eqnarray}
     Further, note that the forward measurement operators satisfy the completeness relation,
     \begin{equation}
     \hat{M}_{F}^{\dagger}(0)\hat{M}_{F}(0)+\hat{M}_{F}^{\dagger}(1)\hat{M}_{F}(1) = \mathbb{I}.\label{cmplt}
     \end{equation}
     We now derive the time reversed measurement operators in this case through the approach we take in this manuscript. Under time reversal, the c-NOT unitary is modified in the following manner,
     \begin{equation}
     \tilde{U}_{\text{c-NOT}}=\Theta U_{\text{c-NOT}}\Theta^{-1} = \mathbb{I}\otimes|1\rangle\langle 1|-\sigma_{x}\otimes|0\rangle\langle 0|.
     \end{equation}
     This can be verified immediately from the following representation for the anti-unitary time reversal operator $\Theta$ for a qubit, $\Theta\equiv i~\hat{\sigma}_{y}\text{K}$, where $i~\hat{\sigma}_{y}$ is unitary, $\hat{\sigma}_{y}$ being the familiar Pauli matrix observable for spin along $y$ axis, and $\text{K}$ indicating complex conjugation. Also note that under time reversal, a generic spin observable $\hat{n}.\vec{\sigma}$ flips sign,
     \begin{equation}
     \Theta\hat{n}.\vec{\sigma}\Theta^{-1} = -\hat{n}.\vec{\sigma}.
     \end{equation}
     We also obtain the relations, $|\tilde{0}\rangle = \Theta|0\rangle = -|1\rangle$ and $|\tilde{1}\rangle = \Theta|1\rangle = |0\rangle$ using $\Theta\equiv i~\hat{\sigma}_{y}\text{K}$. The time reversed initial state, $|\tilde{\psi}_{m}\rangle\otimes|\tilde{\psi}_{s}\rangle$, is given by,
     \begin{equation}
     |\tilde{\psi}_{m}\rangle\otimes|\tilde{\psi}_{s}\rangle = \Theta |\psi_{m}\rangle\otimes|\psi_{s}\rangle =(-\sqrt{\gamma}|1\rangle +\sqrt{1-\gamma}|0\rangle)\otimes|\tilde{\psi}_{s}\rangle, 
     \end{equation}
     The time reversed final state becomes,\begin{equation}
     |\tilde{\psi}_{f}\rangle
     =\tilde{U}_{\text{c-NOT}}|\tilde{\psi}_{m}\rangle\otimes|\tilde{\psi}_{s}\rangle = (-\sqrt{\gamma}|1\rangle +\sqrt{1-\gamma}|0\rangle)\otimes|1\rangle\langle 1||\tilde{\psi}_{s}\rangle+(\sqrt{\gamma}|0\rangle -\sqrt{1-\gamma}|1\rangle)\otimes|0\rangle\langle 0||\tilde{\psi}_{s}\rangle.
     \end{equation}
     Now we compute the time reversed measurement operators using our prescription, which is to project the final state to the time reversed state of the measuring device corresponding to the readout. We obtain,
     \begin{eqnarray}
     \langle \tilde{0}_{m}|\tilde{\psi}_{f}\rangle &=& -\langle 1_{m} |\tilde{\psi}_{f}\rangle = \big[\sqrt{1-\gamma}|0\rangle\langle 0|+\sqrt{\gamma}|1\rangle\langle 1|\big]|\tilde{\psi}_{s}\rangle \equiv \tilde{M}_{R}(0)|\tilde{\psi}_{s}\rangle,~\text{and}\\
          \langle \tilde{1}_{m}|\tilde{\psi}_{f}\rangle &=& \langle 0_{m} |\tilde{\psi}_{f}\rangle = \big[\sqrt{\gamma}|0\rangle\langle 0|+\sqrt{1-\gamma}|1\rangle\langle 1|\big]|\tilde{\psi}_{s}\rangle \equiv \tilde{M}_{R}(1)|\tilde{\psi}_{s}\rangle.
     \end{eqnarray}  Further, note that the time reversed measurement operators also form the same POVM set satisfying the completeness relation in Eq.~\eqref{cmplt}. The two elements in the set are cofactor matrices of each other, which means that they are proportional to the inverses of each other, where the proportionality constant is their determinant. We make the following assertions as a consequence,
     \begin{equation}
     \tilde{M}_{R}(0) = \Theta\hat{M}_{F}(0)\Theta^{-1}\equiv\hat{M}_{F}(1),\hspace{1cm}\text{and}\hspace{1cm}\tilde{M}_{R}(1) = \Theta\hat{M}_{F}(1)\Theta^{-1}\equiv\hat{M}_{F}(0),
     \end{equation}
     where the time reversal operator exchanges the roles of $0$ and $1$ of the measuring device. This illustrates that our rules for time reversing measurement operators are also applicable when the measuring device is also a qubit. The measurement operators we considered so far, both the Gaussian POVM and the two outcome POVM for measuring the spin, were Hermitian. In the following we consider a different scenario of measurement of the fluorescence of a qubit, where the measurement operators are not Hermitian.  
     \subsection{Fluorescence measurement of a qubit\label{Fl}}
     Here we consider the case of a non-Hermitian measurement operator which appear in the fluorescence measurements of a qubit~\cite{jordan2016anatomy}. The fluorescence can be modeled as a special case of an atom ($A$) interacting with a single mode of a field ($B$), where the field initially is in the vacuum state $(|0\rangle)$ while the atom can be in a generic state $|\psi\rangle = a|0\rangle+b|1\rangle$~\cite{jordan2016anatomy}. Following the resonant interaction for a duration $\delta t$, the atom and the field mode get entangled to the final state~\cite{jordan2016anatomy},
     \begin{equation}
     |\psi_{f}\rangle = b\sqrt{1-\epsilon}|1\rangle_{A}|0\rangle_{B}+b\sqrt{\epsilon}|0\rangle_{A}|1\rangle_{B}+a|0\rangle_{A}|0\rangle_{B}.
     \end{equation}  
     Here $\epsilon=\gamma_{1}\delta t$, where $\gamma_{1}$ is the relaxation rate. Subscripts $A$ and $B$ stand for the two level atom and the bosonic field mode respectively. The fluorescence measurement is completed by projecting the field mode onto a coherent state $|\alpha\rangle$, which determines the forward measurement operators $\hat{M}_{F}(\alpha)$:
     \begin{eqnarray}
     \hat{M}_{F}(\alpha)|\psi\rangle &=& \langle\alpha|\psi_{f}\rangle =e^{-\frac{|\alpha|^{2}}{2}}(b\sqrt{1-\epsilon}|1\rangle+ \alpha^{\ast}b\sqrt{\epsilon}|0\rangle+a|0\rangle)\\
     &=&e^{-\frac{|\alpha|^{2}}{2}}[|0\rangle\langle 0|+\sqrt{1-\epsilon}|1\rangle\langle 1|+\alpha^{\ast}\sqrt{\epsilon}|0\rangle\langle 1|]|\psi\rangle.
     \end{eqnarray}
     The forward measurement operators satisfy the (over)completeness relation:
     \begin{equation}
     \int \frac{d^{2}\alpha}{\pi}\hat{M}_{F}(\alpha)^{\dagger}\hat{M}_{F}(\alpha) = \mathbb{I}.
     \end{equation}
     Now let us consider our microscopic approach to defining time reversed measurement operators. The time reversed final state $|\tilde{\psi}_{f}\rangle$ is,
     \begin{equation}
     |\tilde{\psi}_{f}\rangle=\Theta|\psi_{f}\rangle = b^{\ast}\sqrt{1-\epsilon}|0\rangle_{A}|0\rangle_{B}-b^{\ast}\sqrt{\epsilon}|1\rangle_{A}|1\rangle_{B}-a^{\ast}|1\rangle_{A}|0\rangle_{B},
    \label{cohrev} \end{equation}
   where we used the fact that the ground state and the first excited state of the field are time reversal invariant. Time reversal of a coherent state $|\alpha\rangle$ is the state $|\alpha^{\ast}\rangle$, which follows from the eigenvalue equation that defines a coherent state: $\hat{a}|\alpha\rangle = \alpha|\alpha\rangle$, where $\hat{a}$ is the field annihilation operator. By taking the time reversal of this defining equation, we obtain $\Theta \hat{a}\Theta^{-1}\Theta|\alpha\rangle = \alpha^{\ast}\Theta|\alpha\rangle$. Since $\Theta \hat{a}\Theta^{-1} = \hat{a}$, it follows that $\Theta|\alpha\rangle$ is also an eigenstate of the field annihilation operator $\hat{a}$ with eigenvalue $\alpha^{\ast}$, thereby asserting $\Theta|\alpha\rangle=|\alpha^{\ast}\rangle$. We can also see this from the definition of a coherent state $|\alpha\rangle=e^{\frac{-|\alpha|^{2}}{2}}\sum_{n}\frac{\alpha^{n}}{\sqrt{n!}}|n\rangle$, by applying the time reversal operator directly to the state. 
   By projecting the field mode in Eq.~\eqref{cohrev} onto the time reversed coherent state $\Theta|\alpha\rangle=|\alpha^{\ast}\rangle$, we obtain the time reversed measurement operator, $\tilde{M}_{R}(\alpha)$,
     \begin{eqnarray}
     \tilde{M}_{R}(\alpha)|\tilde{\psi}\rangle &=& \langle\alpha^{\ast}|\tilde{\psi}_{f}\rangle =e^{-\frac{|\alpha|^{2}}{2}}(b^{\ast}\sqrt{1-\epsilon}|0\rangle- \alpha b^{\ast}\sqrt{\epsilon}|1\rangle-a^{\ast}|1\rangle)\\
     &=&e^{-\frac{|\alpha|^{2}}{2}}[|1\rangle\langle 1|+\sqrt{1-\epsilon}|0\rangle\langle 0|-\alpha\sqrt{\epsilon}|1\rangle\langle 0|]|\tilde{\psi}\rangle,
     \end{eqnarray}
     where $|\tilde{\psi}\rangle=\Theta|\psi\rangle = -a^{\ast}|1\rangle+b^{\ast}|0\rangle$. We further make the assertion that,   
     \begin{equation}
     \tilde{M}_{R}(\alpha)\equiv\Theta\hat{M}_{F}(\alpha)\Theta^{-1}.
     \end{equation} 
     The time reversed measurement operators also form a POVM since they satisfy the (over)completeness relation,
     \begin{equation}
     \int \frac{d^{2}\alpha}{\pi}\tilde{M}_{R}(\alpha)^{\dagger}\tilde{M}_{R}(\alpha) = \mathbb{I}.
     \end{equation} 
 The three examples discussed so far were specific, but ubiquitous in experiments using weakly or continuously monitored quantum systems~\cite{vijay2012stabilizing,hacohen2016quantum,weber2014mapping,naghiloo2018information,campagne2014observing,campagne2016observing,naghiloo2017thermodynamics,naghiloo2017quantum}. In the following, we show that our approach to  obtaining measurement operators in the time reversed reference frame is more general than~   the particular illustrative examples we presented, and gives a better understanding of how to  unravel the quantum trajectory of a qubit   subjected to generalized measurements.  
 \subsection{A generic rank two qubit measurement\label{gen}}   
 So far we considered very specific examples to demonstrate time reversal of measurement operators. Here we conclude this discussion by noticing that our approach is quite general and can be used to study time reversal of measurement operators having a microscopic derivation similar to our discussion in the previous sections. To show this, we consider a qubit initialized in an arbitrary state $|\psi\rangle$, and the measuring device in the state $|\psi_{M}\rangle$. After the qubit and the measuring device interact, the joint time evolved state can be written in the following generic form,
 \begin{equation}
 |\psi_{f}\rangle=\big(|\psi_{00}\rangle|0\rangle\langle 0|+|\psi_{01}\rangle|0\rangle\langle 1|+|\psi_{10}\rangle|1\rangle\langle 0|+|\psi_{11}\rangle|1\rangle\langle 1|\big)|\psi\rangle,
 \end{equation}
 in the Schr\"{o}dinger picture. The unnormalized states $|\psi_{ij}\rangle~\epsilon~\mathcal{H}_{M}:$ Hilbert space of the measuring device, where~$i, j~\epsilon~\{0,1\}$, contain information about the interaction. The forward measurement operator can be found by projecting the measuring device onto a particular state $|\phi\rangle$,
 \begin{equation}
 \hat{M}_{F}|\psi\rangle=\langle\phi|\psi_{f}\rangle = \big(\langle\phi|\psi_{00}\rangle|0\rangle\langle 0|+\langle\phi|\psi_{01}\rangle|0\rangle\langle 1|+\langle\phi|\psi_{10}\rangle|1\rangle\langle 0|+\langle\phi|\psi_{11}\rangle|1\rangle\langle 1|\big)|\psi\rangle.
 \end{equation} 
 Now the time reversed measurement operators can be defined by first considering the time reversed final state, $|\tilde{\psi}_{f}\rangle=\Theta|\psi_{f}\rangle$, and projecting the measuring device onto the time reversed state $|\tilde{\phi}_{f}\rangle=\Theta|\phi_{f}\rangle$:
 \begin{equation}
 \tilde{M}_{R}|\tilde{\psi}\rangle=\langle\tilde{\phi}|\tilde{\psi}_{f}\rangle = \big(\langle\tilde{\phi}|\tilde{\psi}_{00}\rangle|1\rangle\langle 1|-\langle\tilde{\phi}|\tilde{\psi}_{01}\rangle|1\rangle\langle 0|-\langle\tilde{\phi}|\tilde{\psi}_{10}\rangle|0\rangle\langle 1|+\langle\tilde{\phi}|\tilde{\psi}_{11}\rangle|0\rangle\langle 0|\big)|\tilde{\psi}\rangle,
 \end{equation}
 and we assert that $\tilde{M}_{R}\equiv\Theta \hat{M}_{F}\Theta^{-1}$ in agreement with our discussions in the previous sections. Since the time reversal operator is anti-unitary, we also have $\langle\tilde{\phi}|\tilde{\psi}_{ij}\rangle=\langle\phi|\psi_{ij}\rangle^{\ast}~\forall~i,j~\epsilon~\{0,1\}$. Now a little algebra leads to the straightforward conclusion for rank two qubit measurement operators we consider, 
 \begin{equation}
 \tilde{M}_{R}=\Theta \hat{M}_{F}\Theta^{-1}=\text{det}\big(\hat{M}_{F}^{\dagger}\big)~\big(\hat{M}_{F}^{\dagger}\big)^{-1},\label{reversing}
 \end{equation}
 where the invertibility of $\hat{M}_{F}^{\dagger}$ is guaranteed since the measurement operators are of rank 2. From Eq.~\eqref{reversing} we notice that $\hat{M}_{F}^{\dagger}\tilde{M}_{R}\propto\mathbb{I}$. The qubit trajectory can be reversed by first applying the time reversal operation to the final state, and then applying the Hermitian conjugate of forward sequence of measurement operators in reverse order, which will take the system back to the time reversed initial state. This property is discussed in detail in the coming section. We also add that while the above relation holds true for any two by two matrix with complex elements, they have to satisfy the completeness relation for a particular POVM set in order to qualify as a valid measurement operator, reminiscent of the fact that they can be derived from a microscopic unitary dynamics. 
 
 Further note that our discussion here is closely related to the discussion of undoing quantum measurements by further measurements~\cite{jordan2010uncollapsing,korotkov2006undoing}; In particular, Eq.~\eqref{reversing} also tells us that the the measurement which undoes a previous measurement will be proportional to the Hermitian conjugate of $\tilde{M}_{R}$, since $\tilde{M}_{R}^{\dagger}\hat{M}_{F}\propto \mathbb{I}$. Such inverse measurements in general can be understood as a combination of unitary rotations and a dispersive measurement, using the singular value decomposition of the measurement operator, discussed in Sec.~\ref{Sec. TR}~\cite{jordan2010uncollapsing}.
   \section{Unraveling the quantum state dynamics\label{Sec. TR}}  
 Recall that the statement of time reversal symmetry of Schr\"{o}dinger equation in quantum mechanics is equivalent to the following statement: A quantum system evolves unitarily for a time $\tau$ from $|\psi(0)\rangle$ to $|\psi(\tau)\rangle$ via the operation U$(\tau)$, and if the generator of unitary dynamics (i.e. Hamiltonian for time evolution) is time reversal invariant, the system evolves to the time reversed initial state $\Theta|\psi(0)\rangle$, by the action of the same unitary operator U$(\tau)$ following the time reversal operation on the final state:
 \begin{equation}
 |\psi(0)\rangle\xrightarrow{\text{U}(\tau)}|\psi(\tau)\rangle\xrightarrow{\Theta}\Theta|\psi(\tau)\rangle\xrightarrow{\text{U}(\tau)}\Theta|\psi(0)\rangle\xrightarrow{\Theta} \eta|\psi(0)\rangle,\label{revg}
 \end{equation} 
 where we also note that time reversal operation is reversible up to a global factor $\eta$ (This follows from the standard form of the time reversal operator, $\Theta\equiv\hat{U}\text{K}:~\text{we have}~\Theta^{2} = \hat{U}\text{K}\hat{U}\text{K}=\hat{U}\hat{U}^{\ast}\text{K}^{2}= \hat{U}\hat{U}^{\ast}=\eta$, where $\eta$ is a complex number. Since $\eta\hat{U}^{\dagger} = \hat{U}^{\ast}$, we have $\hat{U}^{\ast}=\eta \hat{U}^{\dagger}=\eta (\hat{U}^{\ast})^{T} = \eta(\eta \hat{U}^{\dagger})^{T} = \eta^{2}\hat{U}^{\ast}$, showing that $\Theta^{2}=\eta=\pm 1$~\cite{haake2013quantum}). Note that a similar statement to Eq.~\eqref{revg} is true for single qubit unitary operators, although since the generators of unitary rotations in the Bloch sphere flips sign under time reversal, the unitary operator that reverses the dynamics after applying the time reversal operator is the Hermitian conjugate of the original unitary operator. This statement can be extended to the case of generalized measurements using the methods we discussed in Sec.~\ref{gen}. We now propose the following generalized statement  about unraveling the quantum state dynamics subjected to measurements:   
 \begin{equation}
 |\psi(0)\rangle\xrightarrow{\hat{M}_{F}(\bar{r})}\hat{M}_{F}(\bar{r})|\psi(0)\rangle\propto|\psi(\tau)\rangle\xrightarrow{\Theta}\Theta|\psi(\tau)\rangle \propto\tilde{M}_{R}(\bar{r})\Theta|\psi(0)\rangle \xrightarrow{\hat{M}_{F}^{\dagger}(\bar{r})}\hat{M}_{F}^{\dagger}(\bar{r})\tilde{M}_{R}(\bar{r})\Theta|\psi(0)\rangle\propto\Theta|\psi(0)\rangle,\label{rev}
 \end{equation} 
 where $``\propto"$ indicates that the statement is true up to a normalization constant which can be determined. The above statement is valid for any rank two qubit operators, including single qubit unitaries. Note that we are effectively flipping the sign of any Rabi drive in the reverse dynamics since we are applying the Hermitian conjugate of the unitary operator in retrodiction, in agreement with the approach in Ref.~\cite{dressel2017arrow}.  Further applying $\Theta$ at the end of Eq.~\eqref{rev} again reverses the time reversal operator up to a sign, since $\Theta^{2}=-1$ for spin systems. We thus obtain the analogous conclusion, that a qubit subjected to a generalized spin measurement, following time reversal operation, evolves towards the time reversed initial state, when acted upon again by the (Hermitian conjugate of the) measurement operator. We call Eq.~\eqref{rev} the active perspective of time reversal in this manuscript, where the reversed dynamics occur through sign flipped Bloch sphere coordinates. Further note that Eq.~\eqref{reversing} also implies $\tilde{M}_{R}^{\dagger}\hat{M}_{F}\propto \mathbb{I}$. Hence there is an equivalent description for time reversing qubit trajectories without having to sign flip the Bloch sphere coordinates. We call this the passive perspective of time reversal where we do not apply the time reversal operator to the final state, rather apply the (Hermitian conjugate of) time reversed measurement operators in reverse order in order to undo a set of forward measurements. A more detailed description of the two perspectives for Gaussian spin measurements can be found in Ref.~\cite{dressel2017arrow}. We also discuss this in Sec.~\ref{arrow}.
 
 We also note that updating information using Hermitian conjugate of measurement operators is known as retrodiction of the measurement~\cite{tan2015prediction}. In the active perspective of time reversal, the equations of motion for qubit trajectories for the reversed dynamics would be identical to the retrodicted trajectory equations~\cite{tan2015prediction}, while the major difference is that the time reversed dynamics starts from the time reversed final state of the qubit, and retrodiction evolves this state to the time reversed initial state. We stress that this is a stronger result than the time reversal invariance of quantum measurements usually considered (see, for example, Refs.~\cite{aharonov1964time,bedingham2017time}), since our approach also keeps the state dynamics time reversal invariant in addition to the probabilistic equivalence of measurement operators discussed in Appendix.~\ref{equivP}. The general procedure of retrodiction can be applied to strong measurements as well, so one might naively think that our rule can be applied to reverse strong measurements too. This is not possible for the following two reasons. First, the time reversal operation takes a qubit state to the orthogonal quantum state, and hence retrodiction from the time reversed final state annihilates the state. Second, the requirement that the measurement operators should have nonzero eigenvalues for them to be invertible; projection operators for qubit are rank one measurement operators and hence they are not invertible. The limit of strong measurements can be taken nevertheless (for example, the limit $\gamma\rightarrow 1$ in the example we consider in Sec.~\ref{qubitmeter}), leading to an arrow of time that is infinite.  
 
 Further note that any arbitrary $2\times 2$ matrix $\hat{M}$ with complex elements can be written as a product of a unitary matrix $\mathcal{U}$, a diagonal matrix with non-negative elements $\mathcal{D}$ and another unitary operator $\mathcal{V}^{\dagger}$ using the singular value decomposition: $\hat{M}=\mathcal{U}\mathcal{D}\mathcal{V}^{\dagger}.$ This alternatively suggest that any single qubit measurement operator can be decomposed into rotations and weak measurements along the computational basis ($\hat{S}_{z}$ basis) described by the operator $\mathcal{D}$, whose diagonal entries are the singular values of $\hat{M}$. This method of decomposing the measurement operator to unitary rotations and a quantum non demolition measurement has been used previously for the purpose of undoing quantum measurements on a qubit~\cite{jordan2010uncollapsing,korotkov2006undoing}.  The advantage of using singular value decomposition is that it allows us to treat readouts of the measurement along the $z$ axis and the rotation angles as invariant quantities under time reversal, while the axis of rotation are sign flipped; the forward measurement can be decomposed into $\hat{S}_{z}$ measurement readouts and angle of rotations, and the  backward   considers the same set of measurement readouts and angle of rotations in the reverse order. 
 \begin{figure}
 	\includegraphics[scale=0.5]{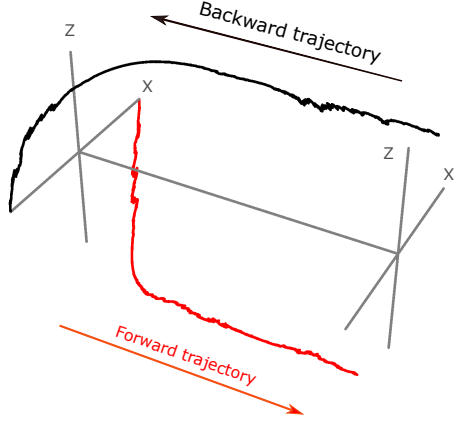}
 	\caption{Here we plot the forward (red) and   backward   (black) Bloch sphere trajectories of a continuously monitored qubit fluorescence experiment starting at $x=1$, where the measurement process is described using the fluorescence measurement operators. The  backward   trajectory obeys the retrodicted equations of motion of the forward trajectory, starting from the time reversed final state and ending at the time reversed initial state, $x=-1$. In the units where $\hbar=1$, we choose the relaxation rate $\gamma_{1}=3.0$, the duration of measurement $T=4.0$, and the Rabi frequency $\Omega=0.5\pi$.\label{fig2}}
 \end{figure}  
 \subsection{Unraveling the fluorescence measurement}  
As a particular example to demonstrate our approach to reversing the measurement dynamics, here we consider the fluorescence measurement of a qubit discussed in Sec.~\ref{Fl}, and show that the  backward   equations correspond to the retrodicted equations for the time reversed quantum state. For the measurement readout $\alpha$, along the unitary time evolution $\hat{U}_{\delta t}$ generated by the qubit Hamiltonian, $\hat{H}=\frac{\Omega}{2}\hat{\sigma}_{y}$, the  forward   time evolution of the density matrix is (units of $\hbar=1$),
\begin{equation}
\rho(t+dt)=\frac{\hat{M}_{F}(\alpha)\hat{U}_{\delta t}\rho(t)\hat{U}_{\delta t}^{\dagger}\hat{M}_{F}^{\dagger}(\alpha)}{\text{Tr}[\hat{M}_{F}(\alpha)\hat{U}_{\delta t}\rho(t)\hat{U}_{\delta t}^{\dagger}\hat{M}_{F}^{\dagger}(\alpha)]}.
\end{equation}
We can also write down the following set of equations for the time evolution of the Bloch coordinates, $x,y$ and $z$~\cite{jordan2016anatomy} for the qubit density matrix $\rho(t)$ defined as $\rho(t)=\frac{1}{2}[\hat{\mathbb{I}}+x(t)\hat{\sigma}_{x}+y(t)\hat{\sigma}_{y}+z(t)\hat{\sigma}_{z}]$:
\begin{eqnarray}
\frac{dx}{dt}&=&\Omega z + \frac{\gamma_{1}}{2}xz+\sqrt{\frac{\gamma_{1}}{2}}[I(1-x^{2}+z)-Qxy]\nonumber\\
\frac{dy}{dt}&=&\frac{\gamma_{1}}{2}yz+\sqrt{\frac{\gamma_{1}}{2}}[Q(1-y^{2}-z)-Ixy]\nonumber\\
\frac{dz}{dt}&=&-\Omega x+\frac{\gamma_{1}}{2}(z^{2}-1)-\sqrt{\frac{\gamma_{1}}{2}}[I(x+xz)+Q(y+yz)].\label{EqFls}
\end{eqnarray}
Note that this is a set of filtering equations given the measurement record $Q$ and $I$. Here the normalized readout variables $Q$ and $I$ are related to $\alpha$ as $Re(\alpha)=I\sqrt{dt/2}$, and $Im(\alpha)=-Q\sqrt{dt/2}$~\cite{jordan2016anatomy}. Now we consider the retrodicted time evolution of a density matrix, $\rho'(t')=\frac{1}{2}[\hat{\mathbb{I}}+x'(t')\hat{\sigma}_{x}+y'(t')\hat{\sigma}_{y}+z'(t')\hat{\sigma}_{z}]$ for the measurement record $\hat{M}_{F}(\alpha)$:
\begin{equation}
\rho'(t'+dt')=\frac{\hat{U}_{\delta t}^{\dagger}\hat{M}_{F}^{\dagger}(\alpha)\rho'(t')\hat{M}_{F}(\alpha)\hat{U}_{\delta t}}{\text{Tr}[\hat{U}_{\delta t}^{\dagger}\hat{M}_{F}^{\dagger}(\alpha)\rho'(t')\hat{M}_{F}(\alpha)\hat{U}_{\delta t}]}.
\end{equation}
One obtains the following equations for the time evolution of the Bloch coordinates, $x',y'$ and $z'$:
\begin{eqnarray}
\frac{dx'}{dt'}&=&-\Omega z'+\frac{\gamma_{1}}{2}x'z'+\sqrt{\frac{\gamma_{1}}{2}}[I(1-x'^{2}-z')-Qx'y']\nonumber\\
\frac{dy'}{dt'}&=&\frac{\gamma_{1}}{2}y'z'+\sqrt{\frac{\gamma_{1}}{2}}[Q(1-y'^{2}+z')-Ix'y']\nonumber\\
\frac{dz'}{dt'}&=&\Omega x'+\frac{\gamma_{1}}{2}(z'^{2}-1)-\sqrt{\frac{\gamma_{1}}{2}}[I(-x'+x'z')+Q(-y'+y'z')].
\end{eqnarray}
We note that if $\rho'(t')$ were a time reversed state, $x'\rightarrow-x,~y'\rightarrow-y$ and $z'\rightarrow-z$, and $t'\rightarrow T-t$, the retrodicted equations are same as the forward equations, but the  time evolution is such that starting from the time reversed final state, the qubit traces the time reversed Bloch sphere coordinates back to the time reversed initial state, consistent with Eq.~\eqref{rev}. Note that in the equations for the time reversed state, we have effectively flipped the sign of $\Omega$ by applying the Hermitian conjugate of the unitary operator in retrodiction. Please see Fig.~\ref{fig2} where we plot the forward and  backward   trajectories for the fluorescence measurement.
   \section{Origin of a statistical arrow of time\label{arrow}}
The main result we have in this article is presented in Eq.~\eqref{rev}, where we showed that given a measurement record, a generalized quantum measurement can be described in both forward/backward directions in time. This time-symmetry manifests in the set of dynamical equations for generalized quantum measurements as the equations being time reversal invariant, thereby demonstrating time reversal invariance in the strongest form. We now look at how one could obtain an arrow of time for generalized measurements, when the equations of motion are time-reversal invariant. A discussion of this can be found in the paper by Dressel et al., using a Bayesian approach~\cite{dressel2017arrow} for the case of Gaussian spin measurements we discussed in Sec.~\ref{SecGaus}. Here we first discuss the essence of their arguments from our point of view, and also extend the notion of a statistical arrow of time to ensembles prepared in a time symmetrical manner using pre- and post-selections. A method to compute the arrow of time for imperfect measurements is discussed in Appendix.~\ref{imperfect}.

Consider a given sequence of measurement outcomes $\{\bar{r}_{n}\}$ where it may be obtained from any rank two qubit measurement process. Our prior knowledge about whether a given sequence corresponds to either the forward or the  backward   protocol is considered to be non-informative (uniform). For a given set of initial and final states $\{|\psi_{i}\rangle,~|\psi_{f}\rangle\}$, where $|\psi_{f}\rangle = \frac{\prod_{n}\hat{M}_{F}(\bar{r}_{n})|\psi_{i}\rangle}{|\prod_{n}\hat{M}_{F}(\bar{r}_{n})|\psi_{i}\rangle|}$, we can compute the probability of obtaining the sequence $\{\bar{r}_{n}\}$ in the forward direction as,
\begin{equation}
\mathcal{P}_{F}(\{\bar{r}_{n}\})=|\prod_{n}\hat{M}_{F}(\bar{r}_{n})|\psi_{i}\rangle|^{2}.
\end{equation}
Similarly, the probability of the  backward   trajectory is given by,
\begin{equation}
\mathcal{P}_{B}(\{\bar{r}_{n'}\})=|\prod_{n'}\hat{M}_{F}^{\dagger}(\bar{r}_{n'})\Theta|\psi_{f}\rangle|^{2}.
\end{equation}
Here $n'$ denotes that the sequence of measurement $\{\bar{r}_{n'}\}$ is considered in the reverse order. The sense in which measurement operators are time reversed by the action of the time reversal operator $\Theta$ is discussed in detail in Sec.~\ref{weak}. For example, Eq.~\eqref{Gsign} describes how the time reversal operator modifies the Gaussian spin measurement operators, where the measurement record is simply sign flipped.  From Eq.~\eqref{rev}, it is clear that,
\begin{equation}
\prod_{n'}\hat{M}_{F}^{\dagger}(\bar{r}_{n'})\Theta|\psi_{f}\rangle\propto\Theta|\psi_{i}\rangle.
\end{equation}
Note that when the measurement operators are Hermitian, reversing the trajectory is equivalent to considering the measurement record in the reverse order, as would be the case for a sequence of measurements described by Gaussian measurement operators. In comparison to the treatment in~\cite{dressel2017arrow}, we add that our approach treats the sequence of measurement readouts $\{\bar{r}_{n}\}$ as invariant for the Gaussian spin measurement. This is called the active perspective of time reversal where only the Bloch sphere coordinates are sign flipped under time reversal: $x(t)\rightarrow -x(T-t)$, $y(t)\rightarrow -y(T-t)$ and $z(t)\rightarrow -z(T-t)$, while the measurement record simply reverses its time ordering~\cite{dressel2017arrow}. In contrast, the measurement record is sign negated in the passive perspective of time reversal for spin measurements: $r(t)\rightarrow -r(T-t)$ under time reversal, while the Bloch sphere coordinates are considered invariant under time reversal~\cite{dressel2017arrow}. If the measurement operators are not Hermitian, one can use singular value decomposition to split the measurement operators into two rotations, and a measurement along $\hat{S}_{z}$ axis, such as the measurement record is still invariant under reversal of the measurement dynamics.  

Assuming uniform prior probabilities $P(F)=P(R)=\frac{1}{2}$, Bayes rule gives the relation~\cite{dressel2017arrow},
\begin{equation}
P(F|\{\bar{r}_{n}\}) = \frac{\mathcal{P}_{F}(\{\bar{r}_{n}\})P(F)}{\mathcal{P}_{F}(\{\bar{r}_{n}\})P(F)+\mathcal{P}_{B}(\{\bar{r}_{n'}\})P(B)} = \frac{\mathcal{R}}{1+\mathcal{R}},
\end{equation}
where $\mathcal{R} = \frac{\mathcal{P}_{F}(\{\bar{r}_{n}\})}{\mathcal{P}_{B}(\{\bar{r}_{n'}\})}$. Dressel et al., defines the quantity $\log{\mathcal{R}}$ to be the statistical arrow of time, as a statistical average over an ensemble of measurement readouts~\cite{dressel2017arrow}.  

\begin{figure}
	\includegraphics[scale=0.5]{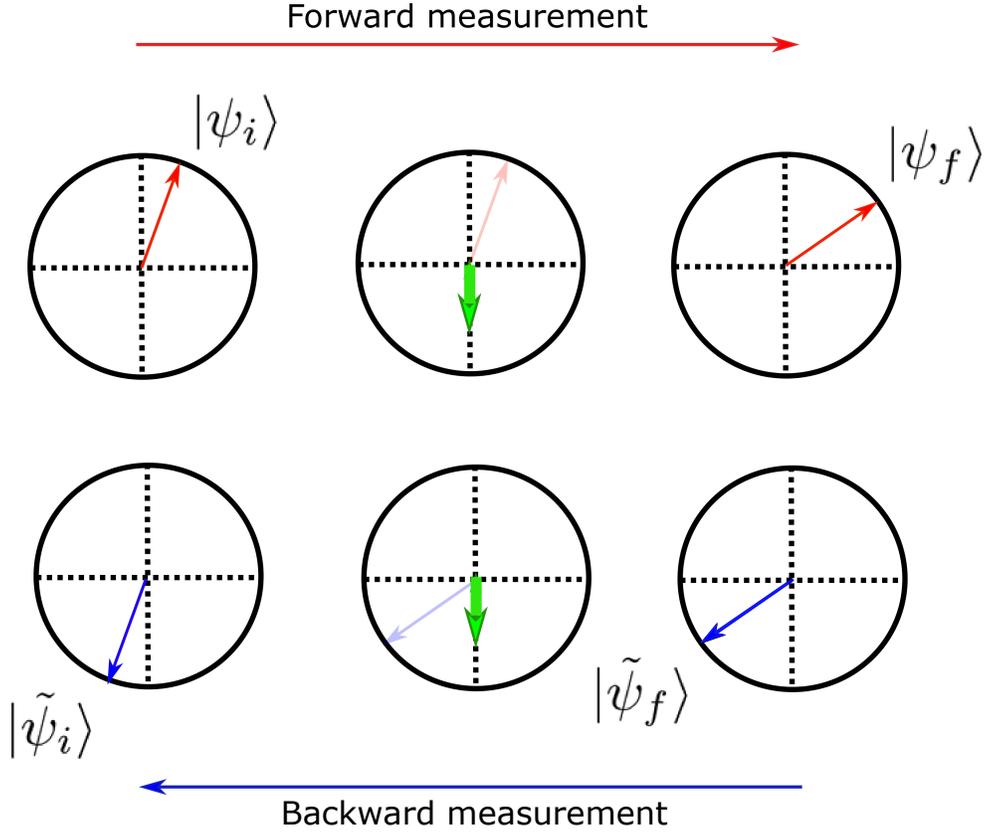}
	\caption{Arrow of time in a continuously monitored qubit measured along the $z$ axis: Here we give a qualitative illustration of an emerging arrow of time, considering a single readout. The long arrows indicate the direction of consideration of events. Consider a qubit state $|\psi_{i}\rangle$ initially pointing closely towards the $|+\rangle$ state (top of Bloch sphere) of the measurement operator $\hat{S}_{z}$. Even though it is highly probable that a weakly coupled measuring device would give a readout corresponding to the $+\frac{\hbar}{2}$ eigenvalue, there is a small nonzero probability that the measurement readout may correspond to the $-\frac{\hbar}{2}$ eigenvalue, indicated by the green arrow. We assume that the measurement was sufficiently weak so the quantum state after the measurement, $|\psi_{f}\rangle$ is still on the upper hemisphere of the Bloch sphere, as shown in the top panel here. Now we consider the  unraveling   of this measurement process, as described using Eq.~\eqref{rev}. We first time reverse the final state, $\Theta|\psi_{f}\rangle=|\tilde{\psi}_{f}\rangle$, upon which the Bloch sphere coordinates of the qubit are parity reversed, and then apply the same measurement operator corresponding to the readout (green arrow), which takes the quantum state back towards the time reversed initial state  $|\tilde{\psi}_{i}\rangle = \Theta|\psi_{i}\rangle$. We notice that for this particular example that we have considered, the time reversed sequence of events are more likely to happen in probability: A quantum state initially pointing towards the state $|+\rangle$ when subjected to a sufficiently weak measurement gives a measurement readout that correspond to the $-\frac{\hbar}{2}$ eigenvalue, taking it more closer towards the $|-\rangle$ eigenstate (bottom of Bloch sphere) of the measurement operator $\hat{S}_{z}$. Since the probability of  the backward   measurement is larger compared to the forward measurement this illustrative example suggests a backward arrow of time.\label{figBWD}}
\end{figure}

The active perspective of time reversal, where the Bloch sphere coordinates are sign flipped, can also be used to make qualitative predictions about when would the arrow of time run backward, like in the case of a continuously monitored qubit measured along the $z$ direction. Because the $z$ coordinate flips sign, a measurement readout $\bar{r}$ is more probable in the backward trajectory when the readout disagrees with the forward measurement outcome in the forward measurement~\cite{dressel2017arrow}. For example, in the forward measurement when the qubit is aligned more towards the positive $z$ direction, statistically it is still possible -- though less probable -- that the readout will correspond to a spin measured along the negative $z$ axis. If the measurement was weak enough, the qubit stays in the upper hemisphere of the Bloch sphere even after the measurement. Under the time reversal operation on the final state, the qubit represented in Bloch sphere coordinates is parity reversed to the lower hemisphere, and then the measurement readout along negative $z$ has a higher probability, indicating an arrow of time that runs backward. This idea is illustrated qualitatively in Fig.~\ref{figBWD}.

In the following, we extend the definition of a statistical arrow of time to pre- and post-selected ensembles and discuss particular examples when the arrow of time would vanish, and an example where it does not vanish. We compute the statistical arrow of time for the fluorescence measurement in Sec.~\ref{FlAoT} and find its stochastic average. A possible extension for the arrow of time measure when the measurements are not perfect is discussed in Appendix.~\ref{imperfect}.
\subsection{Arrow of time in pre- and post-selected sub-ensembles\label{prepost}}
The definition of a statistical arrow of time can also be extended to ensembles prepared in a time symmetric manner via pre- and post-selections onto particular states of the quantum system in the following sense. We again consider a sequence of measurement outcomes $\{\bar{r}_{n}\}$ obtained starting from a pre-selected quantum state $|\phi_{i}\rangle$, and then study the sub-ensemble obtained by post-selecting onto a particular final state $\langle\phi_{f}|$ of the qubit. The forward probability $\mathcal{P}_{F}(\{\bar{r}_{n}\})$ in this case is defined by,
\begin{equation}\label{fw2}
\mathcal{P}_{F}(\{\bar{r}_{n}\}|\phi_{i},\phi_{f}) = |\langle\phi_{f}|\prod_{n}\hat{M}_{F}(\bar{r}_{n})|\phi_{i}\rangle|^2,
\end{equation}
where the forward measurement operators $\hat{M}_{F}(\bar{r}_{n})$ are applied in the order of measurements. The readouts $\bar{r}_{n}$ may also correspond to unitary rotations. The definition of the forward probability, $\mathcal{P}_{F}(\{\bar{r}_{n}\}|\phi_{i},\phi_{f})$, is also in agreement with Ref.~\cite{crooks2008quantum}. Similar to our previous discussions, we may now compute the backward probability, $\mathcal{P}_{B}(\{\bar{r}_{n'}\}|\tilde{\phi}_{i},\tilde{\phi}_{f})$ for the ensemble, by assuming that the sequence of measurements $\{\bar{r}_{n'}\}$ are obtained in the  backward   measurement, starting from the time reversed final state $|\tilde{\phi}_{f}\rangle=\Theta|\phi_{f}\rangle$, apply the same measurement operators in the reverse order [as required from Eq.~\eqref{rev}], and ``post-select" onto the time reversed initial state $\langle\tilde{\phi}_{i}|=\langle\phi_{i}|\Theta$,
\begin{equation}
\label{bw2}
\mathcal{P}_{B}(\{\bar{r}_{n'}\}|\tilde{\phi}_{i},\tilde{\phi}_{f}) = |\langle\tilde{\phi}_{i}|\prod_{n'}\hat{M}_{F}^{\dagger}(\bar{r}_{n'})|\tilde{\phi}_{f}\rangle|^2,
\end{equation} 
where $n'$ denotes that the sequence of measurements are considered in the reverse order. Again, if the forward measurement operators are not Hermitian, one may use the singular value decomposition of the measurement operator such that the measurement record is invariant under time reversal in a certain sense. From a Bayesian point of view, we may quantify this statistical arrow of time as the logarithm of the ratio of the forward and the backward probabilities, $\log{\mathcal{R'}}$, where  $\mathcal{R'} = \frac{\mathcal{P}_{F}(\{\bar{r}_{n}\}|\phi_{i},\phi_{f})}{\mathcal{P}_{B}(\{\bar{r}_{n'}\}|\tilde{\phi}_{i},\tilde{\phi}_{f})}$. We add that although the arrow of time can sensitively depend on the particular pre- and post-selected ensembles, the general statement of the existence of an arrow of time on such sub-ensembles is still valid on more general grounds, since the discussion of a statistical arrow of time of Dressel et al.~\cite{dressel2017arrow}, is a special case of a pre- and post-selected sub-ensemble when the post-selected state coincides with the actual final state obtained at the end of the continuous measurement.

The forward probability $\mathcal{P}_{F}(\{\bar{r}_{n}\}|\phi_{i},\phi_{f}) $ and the backward probability $\mathcal{P}_{B}(\{\bar{r}_{n'}\}|\tilde{\phi}_{i},\tilde{\phi}_{f})$, are in general different for a given set of initial and final states due to the time reversal symmetry operation $\Theta$: the time reversal symmetry operation inverts the parity of the state in the Bloch sphere ($x\rightarrow-x$, $y\rightarrow-y$ and $z\rightarrow-z$), and the probabilities are preserved only in some special cases, as can be seen from a general identity of time reversal operation,
\begin{equation}
\label{timerev1}
\langle\phi|\mathcal{M}|\psi\rangle = \langle\tilde{\psi}|(\Theta \mathcal{M}\Theta^{-1})^{\dagger}|\tilde{\phi}\rangle = \langle\tilde{\psi}| \tilde{\mathcal{M}}^{\dagger}|\tilde{\phi}\rangle.
\end{equation} 
Where $\mathcal{M}$ refers to the product of forward measurement operators appearing in Eq.~\eqref{fw2}. If $\mathcal{M}$ was a two by two unitary matrix, which has the general form $\hat{U}=e^{i\delta}~e^{-i~\hat{n}.\vec{\sigma}\theta}$ for an axis of rotation $\hat{n}$, we note that $\tilde{U}^{\dagger}=e^{2i\delta}\hat{U}^{\dagger}$, and from Equations~\eqref{fw2},~\eqref{bw2}, and~\eqref{timerev1}, we can conclude that the arrow of time vanishes when $\mathcal{M}$ is unitary. Yet another scenario when the arrow of time would vanish can be constructed for pre- and post-selected ensembles. This would correspond to the situation when the pre- and post-selected states are time reversals of each other $\{|\phi_{i}\rangle = |\psi\rangle, |\phi_{f}\rangle = |\tilde{\psi}\rangle\}$, and the product of the sequence measurement operators, $\mathcal{M}$, is either Unitary or Hermitian. To see this we rewrite,  
\begin{equation}\label{ps1}
\mathcal{P}_{F}(\{\bar{r}_{n}\}|\phi_{i},\phi_{f}) = |\langle\phi_{f}|\prod_{n}\hat{M}_{F}(\bar{r}_{n})|\phi_{i}\rangle|^2 = |\langle\tilde{\psi}|\mathcal{M}|\psi\rangle|^2 = \mathcal{M}_{\tilde{\psi},\psi}~(\mathcal{M}^{\dagger})_{\psi,\tilde{\psi}} = |\mathcal{M}_{\tilde{\psi},\psi}|^{2},
\end{equation}
and
\begin{equation}\label{ps2}
\mathcal{P}_{B}(\{\bar{r}_{n'}\}|\tilde{\phi}_{i},\tilde{\phi}_{f}) = |\langle\tilde{\phi}_{i}|\prod_{n'}\hat{M}_{F}^{\dagger}(\bar{r}_{n'})|\tilde{\phi}_{f}\rangle|^2 = |\langle\tilde{\psi}|\mathcal{M}^{\dagger}|\psi\rangle|^2 = (\mathcal{M}^{\dagger})_{\tilde{\psi},\psi}~\mathcal{M}_{\psi,\tilde{\psi}} = |(\mathcal{M}^{\dagger})_{\tilde{\psi},\psi}|^{2}.
\end{equation}
Here we use the notation $\langle\tilde{\psi}|\mathcal{M}|\psi\rangle = \mathcal{M}_{\tilde{\psi},\psi}$, since $|\psi\rangle$ and $|\tilde{\psi}\rangle$ form an orthonormal basis set for the qubit. When the product of the sequence measurement operators $\mathcal{M}$ is Hermitian, we have $\mathcal{M}^{\dagger}=\mathcal{M}$ leading to $\mathcal{P}_{B}(\{\bar{r}_{n'}\}) = \mathcal{P}_{F}(\{\bar{r}_{n}\})$, and a vanishing arrow of time. This would be the case for continuous spin measurements we discussed previously, when the Hamiltonian is set to zero. Further note that if $\mathcal{M}$ is a two by two unitary matrix, it satisfies the relation  $|\mathcal{M}_{\tilde{\psi},\psi}|=|(\mathcal{M}^{\dagger})_{\tilde{\psi},\psi}|$, again leading to a vanishing arrow of time. 

Although particular examples like this can be constructed where the statistical arrow of time vanishes, we emphasize that the conclusion is not generally true in pre and post-selected ensembles, unlike the notion of a vanishing arrow of time that is inbuilt in the ABL rule~\cite{aharonov1964time}; the point of departure from their time symmetry argument is that we consider a microscopic reversal of quantum dynamics which keeps the dynamical equations time reversal invariant; The quantum states in the backward trajectory are either the time reversal of the states we obtained in the forward direction (in the active perspective to time reversal), or they are exactly the same states we obtained in the forward direction (in the passive perspective to time reversal). The ABL rule (or possible extensions to the ABL rule, for example, the one discussed in~\cite{bedingham2017time}) does not discuss such a reversal of the dynamics, rather the time symmetry is similar to what we discussed in Appendix.~\ref{equivP} where the expression for the forward probability can be equally understood as a reverse probability. Our argument is that the arrow of time should not be based on the forward and reverse probabilities defined in Appendix.~\ref{equivP}, rather should arise from considerations of reversal of the quantum dynamics we discussed in Sec.~\ref{Sec. TR} and Sec.~\ref{arrow}.
\subsubsection{Example: non-vanishing arrow of time with past and future boundary conditions}
Here we discuss a simple example to demonstrate the non-vanishing arrow of time with past and future boundary conditions, for the spin measurement process described in Sec.~\ref{qubitmeter}. Consider the situation when we preselect the ensemble to the state $|\psi_{i}\rangle = \alpha_{i}|0\rangle+\beta_{i}|1\rangle$, perform the measurement, and then post-select onto the state  $|\psi_{f}\rangle = \alpha_{f}|0\rangle+\beta_{f}|1\rangle$. When the measuring device collapses to the state $|0_{m}\rangle$, the system being measured is updated by the measurement operator $\hat{M}_{F}(0)$. The forward and backward probabilities can be computed using Eqns.~\eqref{ps1} and~\eqref{ps2}, leading to the statistical arrow of time:
\begin{equation}
\log\mathcal{R} = \log\frac{\mathcal{P}_{F}}{\mathcal{P}_{B}}=2\log\Bigg(\frac{|\sqrt{\gamma}~\alpha_{f}^{\ast}\alpha_{i}+\sqrt{1-\gamma}~\beta_{f}^{\ast}\beta_{i}|}{|\sqrt{1-\gamma}~\alpha_{f}^{\ast}\alpha_{i}+\sqrt{\gamma}~\beta_{f}^{\ast}\beta_{i}|}\Bigg).\label{ex1}
\end{equation}
The measure $\log\mathcal{R}$ is not zero in general. For example, in the above case when the qubit is initialized in the $|\hat{\sigma}_{x},+\rangle$ state with coefficients $\alpha_{i}=\beta_{i}=\frac{1}{\sqrt{2}}$, and after measurement if we post-select the final state to $|\hat{\sigma}_{z},+\rangle$ with coefficients $\alpha_{f}=1,~\beta_{f}=0$, we obtain
$\log\mathcal{R} =\log[\gamma/(1-\gamma)]$. The statistical arrow of time is pointed backward in this example when $\gamma<1/2$. Instead, a post-selection onto the state $|\hat{\sigma}_{z},-\rangle$ with coefficients $\alpha_{f}=0,~\beta_{f}=1$, will give $\log\mathcal{R} =\log[(1-\gamma)/\gamma]$, in which case the arrow of time runs forward when $\gamma<1/2$. We note that the probability of realizing a particular measurement outcome depend sensitively on both the initial and final states and the time ordering of these boundary conditions, leading to a statical arrow of time which is not zero in general. 

However, if we choose $\alpha_{f}=\beta_{i}^{*}$ and $\beta_{f}=-\alpha_{i}^{*}$ in Eq.~\eqref{ex1}, we find that $\log\mathcal{R} = 0$ as an illustration of a vanishing arrow of time when the measurement operator is Hermitian and the pre- and post-selected states are time reversals of each other. Further note that the time symmetric probability relation obtained in Appendix.~\ref{equivP} also extends to situations when we post-select onto particular states of the quantum system being measured. For the example considered above, we find that the reverse probability $p_{R}$ defined in Appendix.~\ref{equivP}, 
\begin{equation}
p_{R} = |\langle\tilde{\psi}_{f}|\tilde{M}_{R}(0)|\tilde{\psi}_{i}\rangle|^{2} = |\sqrt{\gamma}~\alpha_{f}^{\ast}\alpha_{i}+\sqrt{1-\gamma}~\beta_{f}^{\ast}\beta_{i}|^{2}=|\langle\psi_{f}|\hat{M}_{F}(0)|\psi_{i}\rangle|^{2} = p_{F},
\end{equation}
as a consequence of the anti-unitary nature of the time reversal operator $\Theta$. The forward probability $p_{F}\equiv\mathcal{P}_{F}$ by definition, although we use different notations to differentiate between the two contexts discussed in Appendix.~\ref{equivP} and Sec.~\ref{arrow} respectively.  
\subsection{Arrow of time for fluorescence\label{FlAoT}}
In this section, we compute a closed form expression for the statistical arrow of time for the fluorescence measurement we considered in Sec.~\ref{Fl}, and compute its average value, averaged over the noise in the measurement readouts. The probability of obtaining readouts $Q$ and $I$, assuming the measurement is running forward can be written as~\cite{jordan2016anatomy}: \begin{equation}
\mathcal{P}_{F}(\{Q,I\})\propto \exp{\int_{0}^{T}~dt\bigg( \sqrt{\frac{\gamma_{1}}{2}}I(t)x(t) + \sqrt{\frac{\gamma_{1}}{2}}Q(t)y(t) - \gamma_{1}(1+z(t))/2-\frac{I(t)^{2}+Q(t)^{2}}{2}\bigg)},
\end{equation} 
where $\gamma_{1}$ is the relaxation rate and $x(t), y(t)$ and $z(t)$ are the Bloch sphere coordinates as a function of time. The time reversal of this integral is obtained by flipping the sign of the Bloch sphere coordinates at all times. The arrow of time ratio can be obtained as the logarithmic difference between the two:
\begin{equation}
\log\mathcal{R}= 2\int_{0}^{T}dt~\Bigg(\sqrt{\frac{\gamma_{1}}{2}}I(t)x(t) + \sqrt{\frac{\gamma_{1}}{2}}Q(t)y(t)-\gamma_{1}z(t)/2\Bigg).\label{ArFl}
\end{equation}
Interestingly, the last term with coefficient $\gamma_{1}$ indicates that the relaxation process contribute to the arrow of time in addition to the measurement contributions through the first two terms. This is yet another advantage of our prescription for computing the arrow of time, that it is able to capture the contributions to the time's flow due to a physical process inherent to the quantum system, like the atom relaxing to its ground state in this example.
To compute the average value of $\log\mathcal{R}$, we write the Stratonovich integral in Eq.~\eqref{ArFl} as a discrete sum,
 \begin{equation}
 \log\mathcal{R}= 2~dt\sum_{i=1}^{N-1}~\Bigg(\sqrt{\frac{\gamma_{1}}{2}}I_{i}\frac{x_{i}+x_{i+1}}{2} + \sqrt{\frac{\gamma_{1}}{2}}Q_{i}\frac{y_{i}+y_{i+1}}{2}-\gamma_{1}\frac{z_{i}+z_{i+1}}{4}\Bigg).\label{ArFl2}
 \end{equation}
We can write the readouts $I(t)=\sqrt{\frac{\gamma_{1}}{2}}x(t)+\zeta^{x}(t)$ and $Q(t)=\sqrt{\frac{\gamma_{1}}{2}}y(t)+\zeta^{y}(t)$ where $\zeta^{k}:~k=x,y,~$ are Gaussian white noises of variance $\frac{1}{dt}$ each. We find,
\begin{equation}
\overline{\log\mathcal{R}} = \Bigg\langle2~dt\sum_{i=1}^{N-1}~\sqrt{\frac{\gamma_{1}}{2}}\Bigg(\sqrt{\frac{\gamma_{1}}{2}}x_{i}+\zeta^{x}_{i}\Bigg)\frac{2x_{i}+dx_{i}}{2} + \sqrt{\frac{\gamma_{1}}{2}}\Bigg(\sqrt{\frac{\gamma_{1}}{2}}y_{i}+\zeta^{y}_{i}\Bigg)\frac{2y_{i}+dy_{i}}{2}-\frac{\gamma_{1}}{2}\frac{2z_{i}+dz_{i}}{2}\Bigg\rangle,\label{avg}
\end{equation}
where we used the differential form of dynamical equations~\eqref{EqFls}:
\begin{eqnarray}
dx_{i} &=& dt\bigg(\Omega z_{i} + \frac{\gamma_{1}}{2}x_{i}z_{i}+\sqrt{\frac{\gamma_{1}}{2}}\bigg[\bigg(\sqrt{\frac{\gamma_{1}}{2}}x_{i}+\zeta^{x}_{i}\bigg)(1-x_{i}^{2}+z_{i})-\bigg(\sqrt{\frac{\gamma_{1}}{2}}y_{i}+\zeta^{y}_{i}\bigg)x_{i}y_{i}\bigg]\bigg),\nonumber\\\\
dy_{i}&=&dt\bigg(\frac{\gamma_{1}}{2}y_{i}z_{i}+\sqrt{\frac{\gamma_{1}}{2}}\bigg[\bigg(\sqrt{\frac{\gamma_{1}}{2}}y_{i}+\zeta^{y}_{i}\bigg)(1-y_{i}^{2}-z_{i})-\bigg(\sqrt{\frac{\gamma_{1}}{2}}x_{i}+\zeta^{x}_{i}\bigg)x_{i}y_{i}\bigg]\bigg)\nonumber\\\\
dz_{i} &=&dt\bigg(-\Omega x_{i}+\frac{\gamma_{1}}{2}\bigg(z_{i}^{2}-1\bigg)-\sqrt{\frac{\gamma_{1}}{2}}\bigg[\bigg(\sqrt{\frac{\gamma_{1}}{2}}x_{i}+\zeta^{x}_{i}\bigg)(x_{i}+x_{i}z_{i})+\bigg(\sqrt{\frac{\gamma_{1}}{2}}y_{i}+\zeta^{y}_{i}\bigg)(y_{i}+y_{i}z_{i})\bigg]\bigg).
\end{eqnarray}
Substituting this back into Eq~\eqref{avg} and averaging over the noise, we find that terms linear in $\zeta^{k}:~k=x,y,~$ do not contribute to the stochastic average. The quadratic terms  $(\zeta^{k})^{2}:~k=x,y,~$ when multiplied by $dt$ and averaged over the ensemble is equal to one since the variance of $\zeta^{k}:~k=x,y,~$ is $\frac{1}{dt}$. We assume that the two noises corresponding to the readouts $Q$ and $I$ respectively have no cross correlation and hence the ensemble average $\langle\zeta^{x}\zeta^{y}\rangle$ is taken to be zero. We retain terms which are linear in $dt$, and neglect higher order terms leading to the result:
\begin{equation}
\overline{\log\mathcal{R}} \simeq \Bigg\langle 2~dt\sum_{i=1}^{N-1}~\frac{\gamma_{1}}{2}\bigg(x_{i}^{2}+\frac{1-x_{i}^{2}+z_{i}}{2}\bigg) + \frac{\gamma_{1}}{2}\bigg(y_{i}^{2}+\frac{1-y_{i}^{2}-z_{i}}{2}\bigg)-\gamma_{1}\frac{z_{i}}{2}\Bigg\rangle,\label{avg2}
\end{equation}
or in the continuum limit,
\begin{equation}
\overline{\log\mathcal{R}} = \gamma_{1}\int_{0}^{T}dt~\bigg(1+\frac{\langle x^{2}\rangle+\langle y^{2}\rangle}{2}-\langle z\rangle\bigg).\label{ar1}
\end{equation}
where $\langle.\rangle$ indicate that we are averaging over the noise. The last term in Eq.~\eqref{ar1} is interesting. For the fluorescence measurement, the term $-\gamma_{1}\langle z\rangle$ implies that when the qubit spends more time nearer to the ground state ($z=-1$) it speeds up the arrow of time, making it easier to distinguish the times arrow, as opposed to the case when the qubit is close to the excited state ($z=1$) where it slows it down.  

The average value of the arrow of time for the dispersive $z$ measurement has been obtained previously by Dressel et al.~\cite{dressel2017arrow}, as:
\begin{equation}
\overline{\log\mathcal{R}}_{d} = \frac{1}{\tau}\int_{0}^{T}dt~\big(1+\langle z^{2}\rangle\big),\label{ar2}
\end{equation}
where the subscript $d$ indicates dispersive measurement.  We note that the two results in Eq.~\eqref{ar1} and Eq.~\eqref{ar2} are quite similar, where the relaxation time, $\gamma_{1}^{-1}$ in the fluorescence measurement is analogous to the inverse of the measurement strength, $\tau$ for the dispersive measurement.
\section{Special case: Time reversal of the weak value\label{weak2}} 
Weak values in generalized quantum measurements describe anomalous readouts of the measuring device when the system under measurement is post-selected onto a definite final state after the interaction with the measuring device~\cite{aharonov1988result,duck1989sense}. In this section, we discuss what happens to the weak value under time reversal in our approach. To begin with, we again assume that the measuring device is initialized in a zero mean Gaussian quantum state $|\psi_{m}\rangle$, and the unknown quantum state of the system under measurement to be $|\psi_{s}\rangle$, such that the joint initial state of quantum system and the measuring device is given by,
\begin{equation}
|\psi(0)\rangle = |\psi_{m}\rangle\otimes|\psi_{s}\rangle.
\end{equation}
The quantum system and the measuring device evolves unitarily for a duration $\tau$, generated by the interaction Hamiltonian $\hat{H}_{int}  = k~\hat{p}\hat{S}$, 
\begin{equation}
|\psi(\tau)\rangle = e^{-ik\hat{p}\hat{S}\tau}|\psi(0)\rangle=e^{-ik\hat{p}\hat{S}\tau}|\psi_{m}\rangle\otimes|\psi_{s}\rangle=(1-ik\hat{p}\hat{S}\tau+...)|\psi_{m}\rangle\otimes|\psi_{s}\rangle.
\end{equation}
Now the state of the system is post-selected onto a particular final state $\langle\psi_{f}|$, leaving the unobserved state of the measuring device in the following state,
\begin{eqnarray}
\langle\psi_{f}|\psi(\tau)\rangle &=& \langle\psi_{f}|e^{-ik\hat{p}\hat{S}\tau}|\psi_{m}\rangle\otimes|\psi_{s}\rangle=\langle\psi_{f}|(1-ik\hat{p}\hat{S}\tau+...)|\psi_{m}\rangle\otimes|\psi_{s}\rangle\\
&=&(\langle\psi_{f}|\psi_{s}\rangle -i\hat{p}k\tau\langle\psi_{f}|S|\psi_{s}\rangle+...)|\psi_{m}\rangle=\langle\psi_{f}|\psi_{s}\rangle(1-i\hat{p}k\tau~S_{\omega}+...)|\psi_{m}\rangle\\
&=&\frac{\langle\psi_{f}|\psi_{s}\rangle}{(2\pi\delta^{2})^{1/4}}\int_{-\infty}^{\infty}dx~(1-i\hat{p}k\tau~S_{\omega}+...)~e^{-\frac{x^{2}}{4\delta^{2}}}~|x\rangle\\
&\simeq&\frac{\langle\psi_{f}|\psi_{s}\rangle}{(2\pi\delta^{2})^{1/4}}\int_{-\infty}^{\infty}dx'~e^{-\frac{(x'-k\tau S_{\omega})^{2}}{4\delta^{2}}}~|x'\rangle,\label{s1}
\end{eqnarray}
where we note that the wavefunction of the measuring device is an approximate Gaussian in the position basis,
\begin{equation}
\psi_{m}(\bar{x}) = \langle \bar{x}|\frac{\langle\psi_{f}|\psi_{s}\rangle}{(2\pi\delta^{2})^{1/4}}\int_{-\infty}^{\infty}dx~e^{-\frac{(x-k\tau S_{\omega})^{2}}{4\delta^{2}}}~|x\rangle = \frac{\langle\psi_{f}|\psi_{s}\rangle}{(2\pi\delta^{2})^{1/4}}~e^{-\frac{(\bar{x}-k\tau S_{\omega})^{2}}{4\delta^{2}}},
\end{equation}
and we have used the definition of the weak value, $S_{\omega}$~\cite{aharonov1988result,duck1989sense},
\begin{equation}
S_{\omega} = \frac{\langle\psi_{f}|S|\psi_{s}\rangle}{\langle\psi_{f}|\psi_{s}\rangle}.
\end{equation}
The relevant conditions and validity of the approximations made in Eq.~\eqref{s1} is discussed at length in~\cite{duck1989sense}. We now look at what happens to the ``weak value" under time reversal operation using our approach, by which we mean that, operationally, the time reversed final state $|\tilde{\psi}(\tau)\rangle$, is projected onto the the time reversed post-selected state, $\langle\tilde{\psi}_{f}|$.
\begin{eqnarray}
\langle\tilde{\psi}_{f}|\tilde{\psi}(\tau)\rangle &=& \langle\tilde{\psi}_{f}|e^{ik\hat{p}\hat{S}\tau}|\tilde{\psi}_{m}\rangle\otimes|\tilde{\psi}_{s}\rangle=\langle\tilde{\psi}_{f}|(1+ik\hat{p}\hat{S}\tau+...)|\tilde{\psi}_{m}\rangle\otimes|\tilde{\psi}_{s}\rangle\\
&=&(\langle\tilde{\psi}_{f}|\tilde{\psi}_{s}\rangle +i\hat{p}k\tau\langle\tilde{\psi}_{f}|S|\tilde{\psi}_{s}\rangle+...)|\tilde{\psi}_{m}\rangle=\langle\tilde{\psi}_{f}|\tilde{\psi}_{s}\rangle(1+i\hat{p}k\tau~\tilde{S}_{\omega}+...)|\tilde{\psi}_{m}\rangle\\
&=&\frac{\langle\tilde{\psi}_{f}|\tilde{\psi}_{s}\rangle}{(2\pi\delta^{2})^{1/4}}\int_{-\infty}^{\infty}dx~(1+i\hat{p}k\tau~\tilde{S}_{\omega}+...)~e^{-\frac{x^{2}}{4\delta^{2}}}~|x\rangle\\
&\simeq&\frac{\langle\tilde{\psi}_{f}|\tilde{\psi}_{s}\rangle}{(2\pi\delta^{2})^{1/4}}\int_{-\infty}^{\infty}dx'~e^{-\frac{(x'+k\tau \tilde{S}_{\omega})^{2}}{4\delta^{2}}}~|x'\rangle.\label{s2}
\end{eqnarray}
We note that the wavefunction of the measuring device in the position basis is again an approximate Gaussian,
\begin{equation}
\tilde{\psi}_{m}(\bar{x}) = \langle\bar{x}|\frac{\langle\tilde{\psi}_{f}|\tilde{\psi}_{s}\rangle}{(2\pi\delta^{2})^{1/4}}\int_{-\infty}^{\infty}dx'~e^{-\frac{(x'+k\tau \tilde{S}_{\omega})^{2}}{4\delta^{2}}}~|x'\rangle = \frac{\langle\tilde{\psi}_{f}|\tilde{\psi}_{s}\rangle}{(2\pi\delta^{2})^{1/4}}e^{-\frac{(\bar{x}+k\tau \tilde{S}_{\omega})^{2}}{4\delta^{2}}}\label{weakkk},
\end{equation} 
where we have defined the quantity, $\tilde{S}_{\omega},$
\begin{equation}
\tilde{S}_{\omega} = \frac{\langle\tilde{\psi}_{f}|S|\tilde{\psi}_{s}\rangle}{\langle\tilde{\psi}_{f}|\tilde{\psi}_{s}\rangle}.
\end{equation}
Note that time reversal symmetry establishes the following relations,
\begin{equation}
\langle\psi_{f}|S|\psi_{s}\rangle = \langle\tilde{\psi}_{s}|(\Theta S\Theta^{-1})^{\dagger}|\tilde{\psi}_{f}\rangle = -\langle\tilde{\psi}_{s}|S|\tilde{\psi}_{f}\rangle,
\end{equation}
and
\begin{equation}
\langle\tilde{\psi}_{f}|\tilde{\psi}_{s}\rangle = \langle\psi_{s}|\psi_{f}\rangle.
\end{equation}
We thus obtain the relation,
\begin{equation}
\tilde{S}_{\omega} = \frac{\langle\tilde{\psi}_{f}|S|\tilde{\psi}_{s}\rangle}{\langle\tilde{\psi}_{f}|\tilde{\psi}_{s}\rangle} = -\frac{\langle\psi_{s}|S|\psi_{f}\rangle}{\langle\psi_{s}|\psi_{f}\rangle}=-S_{\omega}^{\ast}.\label{s3}
\end{equation}
where $S_{\omega}$ is the forward weak value for measurement of the operator $S$. From equations~\eqref{s1},~\eqref{s2} and \eqref{s3}, we conclude that the weak value is simply complex-conjugated under the time reversal operation, since Eq.~\eqref{weakkk} represents a measuring device state peaked at $S_{\omega}^{\ast}$. This is what one would naively expect since weak value $S_{\omega}$ is a complex number. In addition to that, we note that the wavefunction of the measuring device after post-selection in the time reversed case is related to the time forward case by the relation, $\tilde{\psi}_{m}(\bar{x}) = \psi^{\ast}_{m}(\bar{x})$, in agreement with the time reversal symmetry of the wavefunction in position basis. The complementary scenario when the qubit is coupled to the position of the measuring device is considered in the Appendix.

It is useful to compute the three probabilities we discuss in this manuscript, $p_{F},~p_{R}$ and $p_{B}$ for the measurement protocol described above. Here $p_{F}$ and $p_{R}$ respectively are the forward and reverse probabilities discussed in Appendix.~\ref{equivP}. The probabilities $p_{F}$ and $p_{B}$ respectively are also the forward and backward probabilities we use to compute the statistical arrow of time in Sec.~\ref{prepost}. We first find the relation between  $p_{F}$ and $p_{R}$. We note that the forward probability $p_{F}(\{|\psi_{s}\rangle,~|\psi_{f}\rangle\})$ of post-selecting the qubit onto the state $|\psi_{f}\rangle$ when the qubit (prior to the measurement) is initialized in the state $|\psi_{s}\rangle$ can be written as,
\begin{eqnarray}
p_{F}(\{|\psi_{s}\rangle,~|\psi_{f}\rangle\}) &=&\int_{-\infty}^{\infty}dx~\langle x |\langle\psi_{f}|\psi(\tau)\rangle\langle\psi(\tau)|\psi_{f}\rangle|x\rangle = \int_{-\infty}^{\infty}dx~|\langle x |\langle\psi_{f}|\psi(\tau)\rangle|^{2}\nonumber\\ &=& \int_{-\infty}^{\infty}dx~|\langle x |\langle\tilde{\psi}_{f}|\tilde{\psi}(\tau)\rangle|^{2} = p_{R}(\{|\tilde{\psi}_{s}\rangle,~|\tilde{\psi}_{f}\rangle\}).\label{weakP}
\end{eqnarray}
where $p_{R}(\{|\tilde{\psi}_{s}\rangle,~|\tilde{\psi}_{f}\rangle\})$ is the probability of post-selecting onto the state $|\tilde{\psi}_{f}\rangle$ in the reverse protocol that we used to find the time reversed weak value earlier in this section. We have also used the relation, $\langle x |\langle\psi_{f}|\psi(\tau)\rangle =\langle\tilde{\psi}(\tau)|\tilde{\psi}_{f}\rangle|x\rangle$ and the time reversal invariance of the state $|x\rangle:~\Theta|x\rangle = |x\rangle$. The equivalence in probabilities discussed in Eq.~\eqref{weakP} is similar to what we have discussed in Appendix.~\ref{equivP} for forward and time reversed measurement operators.

Further note that the probability $p_{F}(\{|\psi_{s}\rangle,~|\psi_{f}\rangle\})$ of post-selecting onto the state $|\psi_{f}\rangle$ in the forward direction can be computed exactly in the following manner. Assume that the initial state of the system, $|\psi_{s}\rangle = \alpha_{s}|0\rangle+\beta_{s}|1\rangle$. The state $|\psi_{F}(\tau)\rangle$ of the forward measurement becomes~\cite{duck1989sense,aharonov1988result}:
\begin{equation}
	|\psi_{F}(\tau)\rangle = \alpha_{s}|\psi_{0}\rangle|0\rangle+\beta_{s}|\psi_{1}\rangle|1\rangle, 
\end{equation}
where $|\psi_{0}\rangle$ and $|\psi_{1}\rangle$ are real Gaussian states of the measuring device with mean values $k\tau/2$ and $-k\tau/2$ respectively (units in which $\hbar=1$). The probability of post-selecting onto the final state $|\psi_{f}\rangle = \alpha_{f}|0\rangle+\beta_{f}|1\rangle$ of the qubit is,
\begin{equation}
p_{F}(\{|\psi_{s}\rangle,~|\psi_{f}\rangle\}) = |\alpha_{f}^{\ast}\alpha_{s}|\psi_{0}\rangle+\beta_{f}^{\ast}\beta_{s}|\psi_{1}\rangle|^{2} = |\varepsilon_{0}|\psi_{0}\rangle+\varepsilon_{1}|\psi_{1}\rangle|^{2},
\end{equation}
where $\varepsilon_{0} = \alpha_{f}^{\ast}\alpha_{s}$ and $\varepsilon_{1}=\beta_{f}^{\ast}\beta_{s}$. We now define the backward probability $p_{B}(\{|\tilde{\psi}_{s}\rangle,~|\tilde{\psi}_{f}\rangle\})$ as the probability of the qubit being post-selected onto (the time reversed initial state in the forward measurement)   $|\tilde{\psi}_{s}\rangle= -\alpha_{s}^{\ast}|1\rangle+\beta_{s}^{\ast}|0\rangle$ when it was initialized in (the time reversed final state in the forward measurement) $|\tilde{\psi}_{f}\rangle = -\alpha_{f}^{\ast}|1\rangle+\beta_{f}^{\ast}|0\rangle$. Since the interaction Hamiltonian is invariant under time reversal operation, we obtain, 
\begin{equation}
|\psi_{B}(\tau)\rangle = \beta_{f}^{\ast}|\psi_{0}\rangle|0\rangle-\alpha_{f}^{*}|\psi_{1}\rangle|1\rangle, 
\end{equation}
and the backward probability,
\begin{equation}
p_{B}(\{|\tilde{\psi}_{s}\rangle,~|\tilde{\psi}_{f}\rangle\}) = |\alpha_{f}^{\ast}\alpha_{s}|\psi_{1}\rangle+\beta_{f}^{\ast}\beta_{s}|\psi_{0}\rangle|^{2} = |\varepsilon_{1}|\psi_{0}\rangle+\varepsilon_{0}|\psi_{1}\rangle|^{2}.
\end{equation}
Since $|\psi_{0}\rangle$ and $|\psi_{1}\rangle$ are real Gaussian quantum states of the measuring device, we have $\langle\psi_{0}|\psi_{1}\rangle = \langle\psi_{1}|\psi_{0}\rangle$. This leads to the result $p_{F}=p_{B}$ and a vanishing arrow of time with past and future boundary conditions for the qubit in this example. To conclude, we find that the three probabilities satisfy the relation $p_{F}=p_{R}=p_{B}$ in the measurement protocol that originally defined the weak value in quantum mechanics~\cite{aharonov1988result,duck1989sense}.
 \section{Conclusions}
 This paper extended the notion of microscopic time reversal in quantum mechanics to include reversible measurements on a qubit described by rank two measurement operators. We derived time reversed measurement operators for generic forward measurements by incorporating Wigner's ideas of time reversal to the measurement model and showed that   the unraveling of the measurement backward is described by the same set of equations as the forward dynamics, starting from the time reversed final state. Particular examples of generalized spin measurements and qubit fluorescence were discussed. We also compared our methods for unraveling quantum measurements to earlier notions of uncollapsing a wavefunction by undoing quantum measurements.
 
 Further, we extended the notion of a statistical arrow of time for qubit measurements to arbitrary rank two qubit measurements, and to ensembles prepared with past and future boundary conditions via pre- and post-selections. We noticed that the backward probabilities can be computed by a process similar to retrodiction from the time reversed final state, and that the statistical arrow of time does not vanish in general situations involving pre- and post-selections; This is remarkably different from the notion of a vanishing arrow of time in quantum mechanics with past and future boundary conditions, where time symmetric probabilities are obtained using the ABL rule. The difference arises from our also demanding time reversal symmetry of the quantum state dynamics, rather than just time reversal symmetry in the probability of measurement results. Sufficient conditions for vanishing time's arrow are also discussed. Our arrow of time is also able to capture the contributions to the time's arrow due to natural physical processes like relaxation of an atom to its ground state. We also revisited the notion of a weak value in quantum mechanics and recovered correct transformation rules for the weak value and the wavefunction of the measuring device under time reversal using our approach. 
 \section*{Acknowledgments}
 We thank Kater Murch and Cyril Elouard for fruitful discussions and suggestions. This work was supported by the John Templeton Foundation grant ID 58558,  the US Army Research Office grant No. W911NF-15-1-0496, the National Science
 Foundation grant DMR-1506081 and the US Department of Energy grant No. DE-SC0017890.
 \appendix
 \section{Observation of momentum of the measuring device\label{appA}} 
We consider a slightly different version of our protocol considered in Sec.~\ref{SecGaus}; Here we assume that the position of the measuring device is coupled to the spin, and we are reading out the momentum of the measuring device. The interaction Hamiltonian in this case is,
 \begin{equation}
 \hat{H}_{int}  = -k~\hat{x}\hat{S}.
 \end{equation}
 Note that since the operator $\hat{x}$ do not change sign under time reversal, the Hamiltonian $\hat{H}_{int} $ is not time reversal invariant. Nevertheless the action of time reversal operation is manifest in a simple way, under which the Hamiltonian flips its sign. We first derive the forward measurement operators for the measurement.
   \subsection{Forward measurement operators for observation of momentum of the measuring device}
 We again assume that the measuring device is initialized in a Gaussian state,
 \begin{equation}
 |\psi_{m}\rangle = \frac{1}{(2\pi\delta^{2})^{1/4}}\int_{-\infty}^{\infty}dp~e^{-\frac{p^{2}}{4\delta^{2}}}~|p\rangle.
 \end{equation}
 We also assume that the unknown quantum state of the system under measurement can be expressed in the eigen basis of the observable $\hat{S}$,
 \begin{equation}
 |\psi_{s}\rangle = \sum_{n}c_{n}|s_{n}\rangle.
 \end{equation}
 The joint initial state of quantum system and the measuring device is given by,
 \begin{equation}
 |\psi(0)\rangle = |\psi_{m}\rangle\otimes|\psi_{s}\rangle.
 \end{equation}
 The quantum system and the measuring device evolves unitarily for a duration $\tau$, generated by the interaction Hamiltonian $\hat{H}_{int} $, 
 \begin{eqnarray}
 |\psi(\tau)\rangle &=& e^{ik\hat{x}\hat{S}\tau}|\psi(0)\rangle =\sum_{n'} \frac{1}{(2\pi\delta^{2})^{1/4}}\int_{-\infty}^{\infty}dp~e^{-\frac{p^{2}}{4\delta^{2}}}~e^{ik\hat{x}s_{n'}\tau}|p\rangle\otimes |s_{n'}\rangle\langle s_{n'}||\psi_{s}\rangle\\
 &=& \sum_{n'} \frac{1}{(2\pi\delta^{2})^{1/4}}\int_{-\infty}^{\infty}dp~e^{-\frac{p^{2}}{4\delta^{2}}}|p+k\tau s_{n'}\rangle\otimes |s_{n'}\rangle\langle s_{n'}||\psi_{s}\rangle
 \\&=&\sum_{n'} \frac{1}{(2\pi\delta^{2})^{1/4}}\int_{-\infty}^{\infty}dp'~e^{-\frac{(p'-k\tau s_{n'})^{2}}{4\delta^{2}}}|p'\rangle\otimes |s_{n'}\rangle\langle s_{n'}||\psi_{s}\rangle\\
 &=& \frac{1}{(2\pi\delta^{2})^{1/4}}\int_{-\infty}^{\infty}dp'~e^{-\frac{(p'-k\tau \hat{S})^{2}}{4\delta^{2}}}|p'\rangle\otimes |\psi_{s}\rangle.
 \label{fwdA}
 \end{eqnarray}
 The generalized measurement corresponds to looking at the momentum of the measuring device, or equivalently projecting to one of the momentum eigenstates $|\bar{p}\rangle$. This gives the forward measurement operators, $\hat{M}_{F}(\bar{p})$:
  \begin{equation}
  \langle\bar{p}|\psi(\tau)\rangle = \frac{1}{(2\pi\delta^{2})^{1/4}}e^{-\frac{(\bar{p}-k\tau \hat{S})^{2}}{4\delta^{2}}}|\psi_{s}\rangle = \hat{M}_{F}(\bar{p})|\psi_{s}\rangle,
  \end{equation}  
  where
  \begin{equation}
  \hat{M}_{F}(\bar{p}) = \frac{1}{(2\pi\delta^{2})^{1/4}}e^{-\frac{(\bar{p}-k\tau \hat{S})^{2}}{4\delta^{2}}}.
  \end{equation}
  \subsection{Time reversed measurement operators for observation of momentum of the measuring device}
  We again demand that  the measurement in the time reversed frame of reference is equivalent to   projecting the time reversed final state to a particular measuring device state $|\tilde{\bar{p}}\rangle = \Theta|\bar{p}\rangle = |-\bar{p}\rangle$, corresponding to the readout $\bar{p}$. Since the Hamiltonian flips sign under time reversal, the time evolution operator retains its form under time reversal. The time reversed final state $|\tilde{\psi}(\tau)\rangle$ is, 
  \begin{eqnarray}
  |\tilde{\psi}(\tau)\rangle&=&\sum_{n'} \frac{1}{(2\pi\delta^{2})^{1/4}}\int_{-\infty}^{\infty}dp~e^{-\frac{p^{2}}{4\delta^{2}}}~e^{ik\hat{x}s_{n'}\tau}|-p\rangle\otimes |s_{n'}\rangle\langle s_{n'}|\Theta|\psi_{s}\rangle\\
  &=& \sum_{n'} \frac{1}{(2\pi\delta^{2})^{1/4}}\int_{-\infty}^{\infty}dp~e^{-\frac{p^{2}}{4\delta^{2}}}|-p+k\tau s_{n'}\rangle\otimes |s_{n'}\rangle\langle s_{n'}|\Theta|\psi_{s}\rangle
  \\&=&\sum_{n'}
  \frac{1}{(2\pi\delta^{2})^{1/4}}\int_{-\infty}^{\infty}dp'~e^{-\frac{(p'-k\tau s_{n'})^{2}}{4\delta^{2}}}|p'\rangle\otimes |s_{n'}\rangle\langle s_{n'}|\Theta|\psi_{s}\rangle\\
  &=&\frac{1}{(2\pi\delta^{2})^{1/4}}\int_{-\infty}^{\infty}dp'~e^{-\frac{(p'-k\tau \hat{S})^{2}}{4\delta^{2}}}|p'\rangle\otimes \Theta|\psi_{s}\rangle.
  \end{eqnarray}
  By projecting to the measuring device state $\langle\tilde{\bar{p}}|=\langle-\bar{p}|$, we obtain,
   \begin{equation}
   \langle-\bar{p}|\tilde{\psi}(\tau)\rangle = \frac{1}{(2\pi\delta^{2})^{1/4}}e^{-\frac{(\bar{p}+k\tau \hat{S})^{2}}{4\delta^{2}}}|\tilde{\psi}_{s}\rangle = \tilde{M}_{R}(\bar{p})|\tilde{\psi}_{s}\rangle,
   \end{equation}  
   where
   \begin{equation}
   \tilde{M}_{R}(\bar{p}) = \frac{1}{(2\pi\delta^{2})^{1/4}}e^{-\frac{(\bar{p}+k\tau \hat{S})^{2}}{4\delta^{2}}}.
   \end{equation}
   We obtain the same relation between the forward and time reversed measurement operators,
   \begin{equation}
   \tilde{M}_{R}(\bar{p}) = \hat{M}_{F}(-\bar{p}),
   \end{equation}
   for the case in which the measurement reading is the momentum of the measuring device. The time reversed measurement operators also form a POVM set again, satisfying the completeness relation,
   \begin{equation}
   \int_{-\infty}^{\infty}d\bar{p}~\tilde {M}_{R}^{\dagger}(\bar{p})\tilde{M}_{R}(\bar{p}) = \hat{\mathbb{I}}.
   \end{equation}
       \section{Probabilistic equivalence of measurement operators under time reversal operation\label{equivP}} 
    The motive of this section is to show that our approach to defining time reversed measurement operators can be naturally associated to time symmetric probabilities of measurement outcomes in a certain sense, analogous to Crooks' method where time symmetric probabilities about an asymptotic equilibrium state is the defining equation for time reversed measurement operators~\cite{crooks2008quantum}.  We stress that our conclusion applies to general quantum systems, and not just qubits.   We begin by noting that the probability of obtaining a readout $\bar{r}$ in the forward direction is given by,
    \begin{equation}
    p_{F}(\bar{r}) = |\langle\bar{r}|\psi(\tau)\rangle|^2=|\hat{M}_{F}(\bar{r})|\psi_{s}\rangle|^{2},
    \end{equation}
    where $|\bar{r}\rangle$ corresponds to any quantum state the measuring device is projected onto in the measurement model and $|\psi(\tau)\rangle$ is the joint state of the quantum system and the measuring device. The time reversals of measurement operators are defined in our approach by projecting the joint final state of the quantum system and the measuring device to the time reversed quantum state of the measuring device corresponding to the readout $\bar{r}$, which is the state $|\tilde{\bar{r}}\rangle = \Theta|\bar{r}\rangle$. The probability of obtaining the result $\tilde{\bar{r}}$ in the time reversed protocol can be computed,
    \begin{equation}
    p_{R}(\tilde{\bar{r}}) = |\langle\tilde{\bar{r}}|\tilde{\psi}(\tau)\rangle|^2=|\tilde{M}_{R}(\bar{r})~|\tilde{\psi_{s}}\rangle|^{2},
    \end{equation}
    where we have also used the definition, $|\tilde{\psi}(\tau)\rangle=\Theta|\psi(\tau)\rangle$. We note that the two probabilities are related in a simple manner as follows. To compute the probabilities, it is required to trace over the Hilbert space of the system in both the cases. We re-write,
    \begin{equation}
    p_{F}(\bar{r})=\sum_{n}\langle\bar{r}|\langle s_{n}||\psi(\tau)\rangle\langle\psi(\tau)||\bar{r}\rangle|s_{n}\rangle = \sum_{n}\langle\psi(\tau)||\bar{r}\rangle|s_{n}\rangle\langle\bar{r}|\langle s_{n}||\psi(\tau)\rangle.\label{pf}
    \end{equation}
    From time reversal symmetry, we note that,
    \begin{equation}
    \langle\psi(\tau)||\bar{r}\rangle|s_{n}\rangle = \langle\tilde{\bar{r}}|\langle \tilde{s_{n}}||\tilde{\psi}(\tau)\rangle,~~~\text{and}~~~\langle\bar{r}|\langle s_{n}||\psi(\tau)\rangle = \langle\tilde{\psi}(\tau)||\tilde{\bar{r}}\rangle|\tilde{s_{n}}\rangle.\label{Eq31}
    \end{equation}
     This follows from the anti-unitary nature of the time reversal operator $\Theta$: We can always represent $\Theta=\hat{U}\text{K}$ where $\hat{U}$ is some unitary operator and $\text{K}$ is the complex conjugation operation. Note that $|\tilde{\psi}\rangle=\Theta|\psi\rangle=\hat{U}\text{K}|\psi\rangle=\hat{U}|\psi^{\ast}\rangle$, and $|\tilde{\phi}\rangle=\hat{U}|\phi^{\ast}\rangle$, leading to the result, $\langle\tilde{\phi}|\tilde{\psi}\rangle=\langle\phi^{\ast}|\hat{U}^{\dagger}\hat{U}|\psi^{\ast}\rangle=\langle\phi^{\ast}|\psi^{\ast}\rangle=\langle\phi|\psi\rangle^{\ast} = \langle\psi|\phi\rangle$, which can also be interpreted as the probability amplitude for starting in the state $|\phi\rangle$ and ending in the state $|\psi\rangle$~\cite{sakurai2017modern,gottfried2013quantum}.   This also means that if we include a post-selection to one of the system states after the measurement, the probability that we get the desired outcome (say $|s_{n}\rangle$) in the forward measurement would be the same as the probability of obtaining its time reversal ($|\tilde{s}_{n}\rangle$)   in the measurement in the time reversed frame of reference.  	 We substitute Eq.~\eqref{Eq31} back in to Eq.~\eqref{pf} and obtain,
    \begin{equation}
    p_{F}(\bar{r})= \sum_{n}\langle\psi(\tau)||\bar{r}\rangle|s_{n}\rangle\langle\bar{r}|\langle s_{n}||\psi(\tau)\rangle = \sum_{n}\langle\tilde{\bar{r}}|\langle \tilde{s_{n}}||\tilde{\psi}(\tau)\rangle\langle\tilde{\psi}(\tau)||\tilde{\bar{r}}\rangle|\tilde{s_{n}}\rangle = p_{R}(\tilde{\bar{r}}).\label{prb2}
    \end{equation}
    The result indicates that our definition of time reversed measurement operators naturally satisfies a time symmetric relation to the forward measurement operators. This time reversal invariance is a consequence of the microscopic time reversal invariance of quantum mechanics, and our method can be contrasted with the technique of Crooks~\cite{crooks2008quantum}, which presupposes the time reversal invariance as a requirement to derive the time reversed measurement operators.   
   \section{Arrow of time for imperfect measurements\label{imperfect}}
   Here we address how to incorporate the measurement inefficiency to the arrow of time analysis of a continuously monitored qubit. The purity of a qubit can be reduced in the process of measurement due to imperfect measurements, for example, when the observer is uncertain about the measurement readout and the quantum state $\rho$ has to be updated as an average over a range of possible readouts~\cite{jacobs2014quantum},
   \begin{equation}\label{noninf}
   \rho_{r}=\frac{1}{\mathcal{P}_{F}(r)}\sum_{s}M_{rs}\rho M_{rs}^{\dagger}.
   \end{equation} 
   While the forward probability $\mathcal{P}_{F}(r)$ in this case is straight forward to compute,
   \begin{equation}
   \mathcal{P}_{F}(r) = \sum_{s}\text{tr}[M_{rs}\rho M_{rs}^{\dagger}], 
   \end{equation}
   The decrease in purity can be considered to be some sort of information loss into an undetermined environment which the observer do not have access to, and hence the reversal of this process, where the purity is increased deterministically by the same amount has to be treated specially. Consider a series of such measurements giving imperfect measurement readouts $\{r,r'..\}$ starting from an initially pure quantum state $\rho$; The final quantum state is,
   \begin{eqnarray}
   \rho_{r,r'..} &=& \frac{\sum_{s,s'..}..M_{r's'}M_{rs}\rho M_{rs}^{\dagger}M_{r's'}^{\dagger}..}{\sum_{s,s'..}\text{tr}[..M_{r's'}M_{rs}\rho M_{rs}^{\dagger}M_{r's'}^{\dagger}..]}=\frac{\sum_{s,s'..}p(\{s,s'..\},\{r,r'..\})\rho_{(\{s,s'..\},\{r,r'..\})}}{p(r,r'..)}\\
   &=&\sum_{s,s'..}P(\{s,s'..\},\{r,r'..\})\rho_{(\{s,s'..\},\{r,r'..\})}.\label{ens}
   \end{eqnarray}
   The final quantum state $\rho_{r,r'..}$ obtained after a sequence of imperfect measurements is a generic mixture of pure states $\rho_{(\{s,s'..\},\{r,r'..\})}=\frac{..M_{r's'}M_{rs}\rho M_{rs}^{\dagger}M_{r's'}^{\dagger}..}{\text{tr}[..M_{r's'}M_{rs}\rho M_{rs}^{\dagger}M_{r's'}^{\dagger}..]}$, sampled according to the distribution, $P(\{s,s'..\},\{r,r'..\})=\frac{p(\{s,s'..\},\{r,r'..\})}{p(r,r'..)}$. Here $p(r,r'..)$ is the marginal probability distribution,
   \begin{equation}
   p(r,r'..) = \sum_{s,s'..}\text{tr}[..M_{r's'}M_{rs}\rho M_{rs}^{\dagger}M_{r's'}^{\dagger}..] =  \sum_{s,s'..}p(\{s,s'..\},\{r,r'..\}),
   \end{equation}
   obtained by summing over the $s$ variables. The generalization to cases when the initial state $\rho$ is mixed is similar and straightforward.  The probability of obtaining the sequence of imperfect measurement readouts $\{r,r'..\}$ in the time forward direction is a sum of forward probabilities of all possible trajectories:
   \begin{equation}
   \mathcal{P}_{F}(\{r,r'..\})= \sum_{s,s'..}\text{tr}[..M_{r's'}M_{rs}\rho M_{rs}^{\dagger}M_{r's'}^{\dagger}..]=\sum_{s,s'..}p(\{s,s'..\},\{r,r'..\}).\label{imf}
   \end{equation}
   Now comes the interesting question of how to compute the backward probability, $\mathcal{P}_{B}(\{r,r'..\})$ for this process, which is required to define the statistical arrow of time. Physically, one would expect that  the unraveling of the measurement dynamics   takes the (time reversed) final mixed state to the (time reversed) initial pure state, and we are interested in computing the probability of this process in a sensible way. 
   
   To achieve this, we note that the  final mixed state in Eq.\eqref{ens} has the operational meaning that it describes an ensemble of pure states with well defined measurement histories corresponding to realizations of all possible sequence of measurement readouts with certain frequency, but when the observer is ignorant. Previously discussed time reversal techniques apply to each of the trajectories individually, taking each member of the ensemble back to the (time reversed) initial pure state. We can use that input to compute the probability of the  backward   process $\mathcal{P}_{B}$, by considering the probabilities of time reversal of individual trajectories, weighted by the probability of choosing a particular (time reversed) state from the ensemble:
   \begin{equation}
   \mathcal{P}_{B}(\{r,r'..\}) = \sum_{s,s'..}\mathcal{P}[b|\tilde{\rho}_{(\{s,s'..\},\{r,r'..\})}]p[\tilde{\rho}_{(\{s,s'..\},\{r,r'..\})}],\label{imb}
   \end{equation}  
   where $p[\tilde{\rho}_{(\{s,s'..\},\{r,r'..\})}] = P(\{s,s'..\},\{r,r'..\})$ is the probability of sampling from different (time reversed) final states $\tilde{\rho}_{(\{s,s'..\},\{r,r'..\})}$ obtained from Eq.~\eqref{ens}, and $\mathcal{P}[b|\tilde{\rho}_{(\{s,s'..\},\{r,r'..\})}]$ is the probability of time reversed trajectories starting from the quantum state $\tilde{\rho}_{(\{s,s'..\},\{r,r'..\})}$:
   \begin{equation}
   \mathcal{P}[b|\tilde{\rho}_{(\{s,s'..\},\{r,r'..\})}]=\text{tr}[M_{rs}^{\dagger}M_{r's'}^{\dagger}..\tilde{\rho}_{(\{s,s'..\},\{r,r'..\})}..M_{r's'}M_{rs}].
   \end{equation}
   Note that Eq.~\eqref{imf} and Eq.~\eqref{imb} are conceptually similar, in that both assume a dynamics one can study in principle and hence reversible, but has to resort to an ensemble description because we do not have all the required information about the measurement dynamics due imperfect measurements, or due to an undetermined environment causing decoherence. Under such circumstances, by defining the backward probability $\mathcal{P}_{B}(\{r,r'..\})$ as in Eq.~\eqref{imb}, we are able to extend the definition of a statistical arrow of time, $\log\mathcal{R}$, where $\mathcal{R}= \frac{\mathcal{P}_{F}(\{r,r'..\})}{\mathcal{P}_{B}(\{r,r'..\})}$ -- to cases when the measurement efficiency is less than one.
      \section{Time reversed weak value for observation of momentum of the measuring device\label{appb}}
   Here we consider the complimentary measurement process where the quantum system is coupled to the measuring device via its position variable and the readout correspond to the conjugate momentum. The interaction Hamiltonian in this case is,
   \begin{equation}
   \hat{H}_{int}  = -k~\hat{x}\hat{S},
   \end{equation}
   which flips sign under the time reversal operation. We first consider the forward case. The measuring device is initialized in a zero mean Gaussian quantum state $|\psi_{m}\rangle$ (in the momentum space), and the the unknown quantum state of the system under measurement is denoted as $|\psi_{s}\rangle$, such
   that the joint initial state of quantum system and the measuring device is given by,
   \begin{equation}
   |\psi(0)\rangle = |\psi_{m}\rangle\otimes|\psi_{s}\rangle.
   \end{equation}
   The quantum system and the measuring device evolves unitarily for a duration $\tau$, generated by the interaction Hamiltonian $\hat{H}_{int} $, 
   \begin{equation}
   |\psi(\tau)\rangle = e^{ik\hat{x}\hat{S}\tau}|\psi(0)\rangle=e^{ik\hat{x}\hat{S}\tau}|\psi_{m}\rangle\otimes|\psi_{s}\rangle=(1+ik\hat{x}\hat{S}\tau+...)|\psi_{m}\rangle\otimes|\psi_{s}\rangle.
   \end{equation}
   Now we post-select the state of the system onto a particular final state $\langle\psi_{f}|$, leaving the unobserved state of the measuring device in the following state,
   \begin{eqnarray}
   \langle\psi_{f}|\psi(\tau)\rangle &=& \langle\psi_{f}|e^{ik\hat{x}\hat{S}\tau}|\psi_{m}\rangle\otimes|\psi_{s}\rangle=\langle\psi_{f}|(1+ik\hat{x}\hat{S}\tau+...)|\psi_{m}\rangle\otimes|\psi_{s}\rangle\\
   &=&(\langle\psi_{f}|\psi_{s}\rangle +i\hat{x}k\tau\langle\psi_{f}|S|\psi_{s}\rangle+...)|\psi_{m}\rangle=\langle\psi_{f}|\psi_{s}\rangle(1+i\hat{x}k\tau~S_{\omega}+...)|\psi_{m}\rangle\\
   &=&\frac{\langle\psi_{f}|\psi_{s}\rangle}{(2\pi\delta^{2})^{1/4}}\int_{-\infty}^{\infty}dp~(1+i\hat{x}k\tau~S_{\omega}+...)~e^{-\frac{p^{2}}{4\delta^{2}}}~|p\rangle\\
   &\simeq&\frac{\langle\psi_{f}|\psi_{s}\rangle}{(2\pi\delta^{2})^{1/4}}\int_{-\infty}^{\infty}dp'~e^{-\frac{(p'-k\tau S_{\omega})^{2}}{4\delta^{2}}}~|p'\rangle,
   \end{eqnarray}
   We note that the wavefunction of the measuring device in the momentum basis is an approximate Gaussian,
   \begin{equation}
   \psi_{m}(\bar{p}) = \langle\bar{p}|\frac{\langle\psi_{f}|\psi_{s}\rangle}{(2\pi\delta^{2})^{1/4}}\int_{-\infty}^{\infty}dp~e^{-\frac{(p'-k\tau S_{\omega})^{2}}{4\delta^{2}}}~|p'\rangle = \frac{\langle\psi_{f}|\psi_{s}\rangle}{(2\pi\delta^{2})^{1/4}}e^{-\frac{(\bar{p}-k\tau S_{\omega})^{2}}{4\delta^{2}}},
   \end{equation}
  and the weak value, $S_{\omega}$ is,
   \begin{equation}
   S_{\omega} = \frac{\langle\psi_{f}|S|\psi_{s}\rangle}{\langle\psi_{f}|\psi_{s}\rangle}.
   \end{equation}
   The same set of operations in the time reversed picture gives the time reversed weak value:
   \begin{eqnarray}
   \langle\tilde{\psi}_{f}|\tilde{\psi}(\tau)\rangle &=& \langle\tilde{\psi}_{f}|e^{ik\hat{x}\hat{S}\tau}|\tilde{\psi}_{m}\rangle\otimes|\tilde{\psi}_{s}\rangle=\langle\tilde{\psi}_{f}|(1+ik\hat{x}\hat{S}\tau+...)|\tilde{\psi}_{m}\rangle\otimes|\tilde{\psi}_{s}\rangle\\
   &=&(\langle\tilde{\psi}_{f}|\tilde{\psi}_{s}\rangle +i\hat{x}k\tau\langle\tilde{\psi}_{f}|S|\tilde{\psi}_{s}\rangle+...)|\tilde{\psi}_{m}\rangle=\langle\tilde{\psi}_{f}|\tilde{\psi}_{s}\rangle(1+i\hat{x}k\tau~\tilde{S}_{\omega}+...)|\tilde{\psi}_{m}\rangle\\
   &=&\frac{\langle\tilde{\psi}_{f}|\tilde{\psi}_{s}\rangle}{(2\pi\delta^{2})^{1/4}}\int_{-\infty}^{\infty}dp~(1+i\hat{x}k\tau~\tilde{S}_{\omega}+...)~e^{-\frac{p^{2}}{4\delta^{2}}}~|-p\rangle\\
   &\simeq&\frac{\langle\tilde{\psi}_{f}|\tilde{\psi}_{s}\rangle}{(2\pi\delta^{2})^{1/4}}\int_{-\infty}^{\infty}dp'~e^{-\frac{(p'-k\tau \tilde{S}_{\omega})^{2}}{4\delta^{2}}}~|p'\rangle,
   \end{eqnarray}
   where we note that the wavefunction of the measuring device in the momentum basis is,
   \begin{equation}
   \tilde{\psi}_{m}(\bar{p}) =  \langle\bar{p}|\frac{\langle\tilde{\psi}_{f}|\tilde{\psi}_{s}\rangle}{(2\pi\delta^{2})^{1/4}}\int_{-\infty}^{\infty}dp'~e^{-\frac{(p'-k\tau \tilde{S}_{\omega})^{2}}{4\delta^{2}}}~|p'\rangle=\frac{\langle\tilde{\psi}_{f}|\tilde{\psi}_{s}\rangle}{(2\pi\delta^{2})^{1/4}}~e^{-\frac{(\bar{p}-k\tau \tilde{S}_{\omega})^{2}}{4\delta^{2}}},
   \end{equation}   
   and the time reversed weak value $\tilde{S}_{\omega}$ is,
   \begin{equation}
   \tilde{S}_{\omega} = \frac{\langle\tilde{\psi}_{f}|S|\tilde{\psi}_{s}\rangle}{\langle\tilde{\psi}_{f}|\tilde{\psi}_{s}\rangle},
   \end{equation} and note that the following relation still holds,
   \begin{equation}
   \tilde{S}_{\omega} = \frac{\langle\tilde{\psi}_{f}|S|\tilde{\psi}_{s}\rangle}{\langle\tilde{\psi}_{f}|\tilde{\psi}_{s}\rangle} = -\frac{\langle\psi_{s}|S|\psi_{f}\rangle}{\langle\psi_{s}|\psi_{f}\rangle}=-S_{\omega}^{\ast}.
   \end{equation}
   where $S_{\omega}$ is the forward weak value for measurement of the operator $S$. We also note that the wavefunction of the measuring device after post-selection in the time reversed case is related to the time forward case by the relation, $\tilde{\psi}_{m}(\bar{p}) = \psi^{\ast}_{m}(-\bar{p})$, which is what one would expect from time reversal symmetry of the wavefunction in the momentum basis.
\bibliographystyle{ieeetr}
\bibliography{ref}

\begin{thebibliography}{10}

\bibitem{shapiro1968electric}
F.~Shapiro, ``Electric dipole moments of elementary particles,'' {\em Soviet
  Physics Uspekhi}, vol.~11, no.~3, p.~345, 1968.

\bibitem{purcell1950possibility}
E.~Purcell and N.~Ramsey, ``On the possibility of electric dipole moments for
  elementary particles and nuclei,'' {\em Physical Review}, vol.~78, no.~6,
  p.~807, 1950.

\bibitem{fortson2003search}
N.~Fortson, P.~Sandars, and S.~Barr, ``The search for a permanent electric
  dipole moment,'' {\em Physics Today}, vol.~56, no.~6, pp.~33--39, 2003.

\bibitem{callen1985thermodynamics}
H.~Callen, {\em Thermodynamics and an Introduction to Thermostatistics}.
\newblock Wiley, 1985.

\bibitem{hawking1985arrow}
S.~W. Hawking, ``Arrow of time in cosmology,'' {\em Physical Review D},
  vol.~32, no.~10, p.~2489, 1985.

\bibitem{zurek2003decoherence}
W.~H. Zurek, ``Decoherence, einselection, and the quantum origins of the
  classical,'' {\em Reviews of modern physics}, vol.~75, no.~3, p.~715, 2003.

\bibitem{maccone2009quantum}
L.~Maccone, ``Quantum solution to the arrow-of-time dilemma,'' {\em Physical
  review letters}, vol.~103, no.~8, p.~080401, 2009.

\bibitem{lebowitz1993boltzmann}
J.~L. Lebowitz, ``Boltzmann's entropy and time's arrow,'' {\em Physics today},
  vol.~46, pp.~32--32, 1993.

\bibitem{struppa2013quantum}
D.~Struppa and J.~Tollaksen, {\em Quantum Theory: A Two-Time Success Story:
  Yakir Aharonov Festschrift}.
\newblock Springer Milan, 2013.

\bibitem{coveney1991arrow}
P.~Coveney and R.~Highfield, ``The arrow of time,'' {\em Nature}, vol.~350,
  no.~6318, pp.~456--456, 1991.

\bibitem{gold1962arrow}
T.~Gold, ``The arrow of time,'' {\em American Journal of Physics}, vol.~30,
  no.~6, pp.~403--410, 1962.

\bibitem{sakurai2017modern}
J.~J. Sakurai and J.~Napolitano, {\em Modern quantum mechanics}.
\newblock Cambridge University Press, 2017.

\bibitem{wigner2012group}
E.~Wigner, {\em Group Theory: And its Application to the Quantum Mechanics of
  Atomic Spectra}.
\newblock Pure and applied physics, Elsevier Science, 2012.

\bibitem{gottfried2013quantum}
K.~Gottfried and T.~Yan, {\em Quantum Mechanics: Fundamentals}.
\newblock Graduate Texts in Contemporary Physics, Springer New York, 2013.

\bibitem{haake2013quantum}
F.~Haake, {\em Quantum signatures of chaos}, vol.~54.
\newblock Springer Science \& Business Media, 2013.

\bibitem{katz2008reversal}
N.~Katz, M.~Neeley, M.~Ansmann, R.~C. Bialczak, M.~Hofheinz, E.~Lucero,
  A.~O'Connell, H.~Wang, A.~Cleland, J.~M. Martinis, {\em et~al.}, ``Reversal
  of the weak measurement of a quantum state in a superconducting phase
  qubit,'' {\em Physical review letters}, vol.~101, no.~20, p.~200401, 2008.

\bibitem{korotkov2006undoing}
A.~N. Korotkov and A.~N. Jordan, ``Undoing a weak quantum measurement of a
  solid-state qubit,'' {\em Physical review letters}, vol.~97, no.~16,
  p.~166805, 2006.

\bibitem{dressel2017arrow}
J.~Dressel, A.~Chantasri, A.~N. Jordan, and A.~N. Korotkov, ``Arrow of time for
  continuous quantum measurement,'' {\em Physical Review Letters}, vol.~119,
  no.~22, p.~220507, 2017.

\bibitem{sudarshan1961stochastic}
E.~Sudarshan, P.~Mathews, and J.~Rau, ``Stochastic dynamics of
  quantum-mechanical systems,'' {\em Physical Review}, vol.~121, no.~3, p.~920,
  1961.

\bibitem{jordan1961dynamical}
T.~F. Jordan and E.~Sudarshan, ``Dynamical mappings of density operators in
  quantum mechanics,'' {\em Journal of Mathematical Physics}, vol.~2, no.~6,
  pp.~772--775, 1961.

\bibitem{kraus1971general}
K.~Kraus, ``General state changes in quantum theory,'' {\em Annals of Physics},
  vol.~64, no.~2, pp.~311--335, 1971.

\bibitem{kraus1983states}
K.~Kraus, {\em States, effects and operations: fundamental notions of quantum
  theory}.
\newblock Springer, 1983.

\bibitem{tan2015prediction}
D.~Tan, S.~Weber, I.~Siddiqi, K.~Moelmer, and K.~Murch, ``Prediction and
  retrodiction for a continuously monitored superconducting qubit,'' {\em
  Physical review letters}, vol.~114, no.~9, p.~090403, 2015.

\bibitem{crooks2008quantum}
G.~E. Crooks, ``Quantum operation time reversal,'' {\em Physical Review A},
  vol.~77, no.~3, p.~034101, 2008.

\bibitem{bedingham2017time}
D.~Bedingham and O.~Maroney, ``Time symmetry in wave-function collapse,'' {\em
  Physical Review A}, vol.~95, no.~4, p.~042103, 2017.

\bibitem{aharonov1964time}
Y.~Aharonov, P.~G. Bergmann, and J.~L. Lebowitz, ``Time symmetry in the quantum
  process of measurement,'' {\em Physical Review}, vol.~134, no.~6B, p.~B1410,
  1964.

\bibitem{bednorz2013noninvasiveness}
A.~Bednorz, K.~Franke, and W.~Belzig, ``Noninvasiveness and time symmetry of
  weak measurements,'' {\em New Journal of Physics}, vol.~15, no.~2, p.~023043,
  2013.

\bibitem{bitbol1986time}
M.~Bitbol, ``Time symmetry and quantum measurements,'' {\em Physics Letters A},
  vol.~115, no.~8, pp.~357--362, 1986.

\bibitem{elouard2017role}
C.~Elouard, D.~A. Herrera-Mart{\'\i}, M.~Clusel, and A.~Auff{\`e}ves, ``The
  role of quantum measurement in stochastic thermodynamics,'' {\em npj Quantum
  Information}, vol.~3, no.~1, p.~9, 2017.

\bibitem{elouard2016stochastic}
C.~Elouard, D.~H. Marti, M.~Clusel, and A.~Auff{\`e}ves, ``Stochastic
  thermodynamics in the quantum regime: From quantum measurement to quantum
  trajectories,'' {\em arXiv preprint arXiv:1603.07266}, 2016.

\bibitem{manzano2017quantum}
G.~Manzano, J.~M. Horowitz, and J.~M. Parrondo, ``Quantum fluctuation theorems
  for arbitrary environments: adiabatic and non-adiabatic entropy production,''
  {\em arXiv preprint arXiv:1710.00054}, 2017.

\bibitem{parrondo2009entropy}
J.~M. Parrondo, C.~Van~den Broeck, and R.~Kawai, ``Entropy production and the
  arrow of time,'' {\em New Journal of Physics}, vol.~11, no.~7, p.~073008,
  2009.

\bibitem{benoist2016entropy}
T.~Benoist, V.~Jaksic, Y.~Pautrat, and C.-A. Pillet, ``On entropy production of
  repeated quantum measurements i. general theory,'' {\em arXiv preprint
  arXiv:1607.00162}, 2016.

\bibitem{aharonov1988result}
Y.~Aharonov, D.~Z. Albert, and L.~Vaidman, ``How the result of a measurement of
  a component of the spin of a spin-1/2 particle can turn out to be 100,'' {\em
  Physical review letters}, vol.~60, no.~14, p.~1351, 1988.

\bibitem{duck1989sense}
I.~Duck, P.~Stevenson, and E.~Sudarshan, ``The sense in which a" weak
  measurement" of a spin-$1/2$ particle's spin component yields a value 100,''
  {\em Physical Review D}, vol.~40, no.~6, p.~2112, 1989.

\bibitem{vijay2012stabilizing}
R.~Vijay, C.~Macklin, D.~Slichter, S.~Weber, K.~Murch, R.~Naik, A.~N. Korotkov,
  and I.~Siddiqi, ``Stabilizing rabi oscillations in a superconducting qubit
  using quantum feedback,'' {\em Nature}, vol.~490, no.~7418, p.~77, 2012.

\bibitem{hacohen2016quantum}
S.~Hacohen-Gourgy, L.~S. Martin, E.~Flurin, V.~V. Ramasesh, K.~B. Whaley, and
  I.~Siddiqi, ``Quantum dynamics of simultaneously measured non-commuting
  observables,'' {\em Nature}, vol.~538, no.~7626, p.~491, 2016.

\bibitem{weber2014mapping}
S.~Weber, A.~Chantasri, J.~Dressel, A.~N. Jordan, K.~Murch, and I.~Siddiqi,
  ``Mapping the optimal route between two quantum states,'' {\em Nature},
  vol.~511, no.~7511, p.~570, 2014.

\bibitem{naghiloo2018information}
M.~Naghiloo, J.~Alonso, A.~Romito, E.~Lutz, and K.~Murch, ``Information gain
  and loss for a quantum maxwell's demon,'' {\em arXiv preprint
  arXiv:1802.07205}, 2018.

\bibitem{jacobs2003project}
K.~Jacobs, ``How to project qubits faster using quantum feedback,'' {\em
  Physical Review A}, vol.~67, no.~3, p.~030301, 2003.

\bibitem{campagne2014observing}
P.~Campagne-Ibarcq, L.~Bretheau, E.~Flurin, A.~Auff{\`e}ves, F.~Mallet, and
  B.~Huard, ``Observing interferences between past and future quantum states in
  resonance fluorescence,'' {\em Physical review letters}, vol.~112, no.~18,
  p.~180402, 2014.

\bibitem{campagne2016observing}
P.~Campagne-Ibarcq, P.~Six, L.~Bretheau, A.~Sarlette, M.~Mirrahimi, P.~Rouchon,
  and B.~Huard, ``Observing quantum state diffusion by heterodyne detection of
  fluorescence,'' {\em Physical Review X}, vol.~6, no.~1, p.~011002, 2016.

\bibitem{naghiloo2017thermodynamics}
M.~Naghiloo, D.~Tan, P.~Harrington, J.~Alonso, E.~Lutz, A.~Romito, and
  K.~Murch, ``Thermodynamics along individual trajectories of a quantum bit,''
  {\em arXiv preprint arXiv:1703.05885}, 2017.

\bibitem{naghiloo2017quantum}
M.~Naghiloo, D.~Tan, P.~Harrington, P.~Lewalle, A.~Jordan, and K.~Murch,
  ``Quantum caustics in resonance-fluorescence trajectories,'' {\em Physical
  Review A}, vol.~96, no.~5, p.~053807, 2017.

\bibitem{barchielli1982model}
A.~Barchielli, L.~Lanz, and G.~Prosperi, ``A model for the macroscopic
  description and continual observations in quantum mechanics,'' {\em Il Nuovo
  Cimento B (1971-1996)}, vol.~72, no.~1, pp.~79--121, 1982.

\bibitem{caves1987quantum}
C.~M. Caves and G.~Milburn, ``Quantum-mechanical model for continuous position
  measurements,'' {\em Physical Review A}, vol.~36, no.~12, p.~5543, 1987.

\bibitem{nielsen2010quantum}
M.~A. Nielsen and I.~Chuang, {\em Quantum computation and quantum information:
  10th Anniversary Edition}.
\newblock Cambridge University Press, 2010.

\bibitem{jordan2016anatomy}
A.~N. Jordan, A.~Chantasri, P.~Rouchon, and B.~Huard, ``Anatomy of
  fluorescence: quantum trajectory statistics from continuously measuring
  spontaneous emission,'' {\em Quantum Studies: Mathematics and Foundations},
  vol.~3, no.~3, pp.~237--263, 2016.

\bibitem{jordan2010uncollapsing}
A.~N. Jordan and A.~N. Korotkov, ``Uncollapsing the wavefunction by undoing
  quantum measurements,'' {\em Contemporary Physics}, vol.~51, no.~2,
  pp.~125--147, 2010.

\bibitem{jacobs2014quantum}
K.~Jacobs, {\em Quantum Measurement Theory and its Applications}.
\newblock Cambridge University Press, 2014.

\end{thebibliography}
   \end{document}